\documentclass[a4paper,11pt]{article}
\usepackage{jheppub} % for details on the use of the package, please
\usepackage[utf8]{inputenc}
\usepackage{graphicx}
\usepackage{amsmath}
\usepackage[compat=1.1.0]{tikz-feynman}
\usepackage{fontenc}
\usepackage{mathtools}
\usepackage{slashed}
\usepackage{amssymb}
\usepackage{tikz-feyn}

\title{Lorentz violating backgrounds from quadratic, shift-symmetric, ultralight dark matter}
\preprint{IPPP/24/21, DESY-24-056}
\author[a]{Minyuan Jiang,}
\author[b]{Benjamin D. Pecjak,}
\author[c]{Gilad Perez,}
\author[c]{Somasundaram Sankaranarayanan}
\affiliation[a]{Deutsches Elektronen-Synchrotron DESY, Notkestr. 85, 22607 Hamburg, Germany}
\affiliation[b]{
Institute for Particle Physics Phenomenology, Durham University, Durham DH1 3LE, UK}
\affiliation[c]{
Department of Particle Physics and Astrophysics, Weizmann Institute of Science,
				Herzl St 234, Rehovot 761001, Israel}

\abstract{We consider an effective theory for a shift-symmetric, quadratically-coupled, ultralight spin-0 field. The leading CP conserving interactions with Standard Model fields in the effective theory arise at dimension 8. We discuss the renormalization group evolution and positivity bounds on these operators, as well as their possible UV origins. Assuming that the spin-0 field is associated with an ultralight dark matter candidate, we discuss the effects of the dimension-8 operators on experiments searching for the oscillation of fundamental constants and Lorentz violation. 
We find that the direct bounds on these two effects are of similar strength but rather weak, corresponding to a UV cutoff scale of keV order, as they are mediated by dimension-8 operators.}
\begin{document}

\maketitle
\section{Introduction}

The possibility that dark matter (DM) consists of  ultralight bosonic particles with sub-eV mass has garnered increasing interest in recent times. The small mass of these particles results in a large occupation number, so that rather than behaving as discrete particles, these particles exhibit collective behavior best described by classical waves.  
A spin-0 ultralight field can generally lead to a viable model of DM via the misalignment mechanism~\cite{Preskill:1982cy, Abbott:1982af, Dine:1982ah}. The field $\phi$ can be CP-odd, such as the QCD axion~~\cite{Peccei:1977hh,Weinberg:1977ma,Wilczek:1977pj,Kim:1979if,Shifman:1979if,Dine:1981rt,Zhitnitsky:1980tq} that was originally proposed to address the strong CP problem, or axion-like particles predicted in other well motivated models of spontaneous U(1) symmetry breaking (see for instance~\cite{Froggatt:1978nt,Gelmini:1980re}). Alternatively, it could be CP-even, as exemplified by dilatons arising from the spontaneous breaking of a conformal sector, or otherwise a pseudo/scalar, ultralight DM (ULDM), hybrid with ill-defined parity transformation, which arises in axion models with more than one source of shift symmetry breaking~\cite{Flacke:2016szy, Banerjee:2018xmn}.

In any case, the specific UV origins of these particles are not our primary concern, as in the present paper we leverage the framework of Effective Field Theory (EFT) to describe their interactions at low energies systematically. Our EFT framework is characterized by two symmetries, which require the invariance of the theory under (1) the shift $\phi \rightarrow \phi+\textbf{const.}$\footnote{Note that this symmetry is softly broken by the mass term.} and (2) a $\mathbb{Z}_2$ transformation $\phi \rightarrow-\phi$.  The shift symmetry protects the DM mass from large corrections and keeps it naturally small, and the $\mathbb{Z}_2$ symmetry implies that $\phi$ couples quadratically to the Standard Model (SM) fields. In addition, we consider only CP-even operators. In fact, in models where the SM fields are not charged under the U(1) axionic symmetry,  we expect generically that the leading interactions between the ULDM and the SM would appear at quadratic order, without the need to impose a $\mathbb{Z}_2$ symmetry, and on top of that one should add that almost all models of an axion-like particle are expected to consist of quadratic coupling to the SM scalar operators~\cite{Banerjee:2022sqg}. 

Our EFT provides a consistent framework for describing Lorentz and in particular Galilean violating backgrounds, which we shall refer to collectively as Lorentz violating (LV). We shall demonstrate that it is comprised of two distinct types of effective operators governing the interactions between the field $\phi$ and SM fields.  One type possesses a scalar structure, leading to the oscillation of fundamental constants such as the electron mass and the fine structure constant, while the other exhibits a tensor structure, resulting in similar signals in experiments searching for the aforementioned
effects, and in addition the violation of the Lorentz symmetry \cite{gupta2022probingpoincareviolation} (in particular the violation of spatial rotation symmetry). Consequently, the EFT links two classes of experiments probing for these different  effects, which are usually considered to be unrelated. Such a theoretical link is timely, as experiments searching for oscillations of fundamental constants by comparing the frequencies in atomic clocks and cavities have undergone significant developments in recent years (see~\cite{Antypas:2022asj} for a review of these experiments), as have experiments focusing on searches for LV effects predicted in theories of quantum gravity (see~\cite{Kostelecky:2008ts} for a summary of the experiments and results). 
Since the experiments we are concerned with (in particular the searches for LV) focus on the Quantum Electrodynamics (QED) sector of the SM, in this work we will focus on the operators built from the $\phi$, electron and photon fields.

This paper is organized as follows. In section~\ref{sec:EFT}, we study the theoretical aspects of the EFT. We first list all the relevant operators and then study the RG flow between them. Additionally, we give an example of potential UV origin of these operators, and explore the constraints imposed on the Wilson coefficients by positivity bounds. In section~\ref{sec:bounds}, we demonstrate how these effective operators affect the aforementioned two types of experiment and derive bounds on the Wilson coefficients. A brief discussion of indirect bounds from astrophysical, cosmological and collider data is presented in section~\ref{sec:others}, and  conclusions are given in section~\ref{sec:dis&con}. 

\section{EFT of ULDM with quadratic coupling to QED}
\label{sec:EFT}
In this section we study the EFT that describes the interaction between a real spin-0 field $\phi$ and the QED sector of the SM. The scalar is assumed to be a singlet under the SM gauge group. As already described, we also assume that the theory is invariant under CP, shift symmetry, and a $\mathbb{Z}_2$, $\phi \rightarrow-\phi$. 

\subsection{Effective operators}
\label{sec:ops}
The EFT is organized as an expansion in $1/\Lambda$, with $\Lambda$ characterizing the UV cut-off of the theory. The effective operators of higher mass dimension are thus suppressed by higher orders of $1/\Lambda$. In our case, the first operator that satisfies all symmetries of the theory arises at dimension seven, and is given by
\begin{equation}
\label{operator7}
\overline{O}_1= \partial_\mu\phi\partial^\mu\phi \, \bar{\psi} \psi \,,
\end{equation}
where $\psi$ stands for a fermion field (for instance, an electron).

At dimension 8 there are four independent CP-even operators~\footnote{Other than these CP-even operators, there are also CP-odd ones that obey the $\mathbb{Z}_2$ and shift symmetries but we do not consider them here.}, in the sense that they are not related to each other by the equations of motion (EoM) or integration by parts. These are 
\begin{align}\label{operators8}
    O_1&=\partial_\mu\phi\partial^\mu\phi \, i \bar{\psi}\gamma^\nu \overset{\leftrightarrow}{D_\nu}\psi , \nonumber\\ 
    O_2 &= \partial_\mu\phi\partial_\nu\phi \, i \bar{\psi}\gamma^\nu\overset{\leftrightarrow}{D^\mu}\psi,\nonumber\\
 O_3 &= \frac{1}{2}\partial_\mu\phi\partial^\mu\phi F_{\nu\rho} F^{\nu\rho} \,, \nonumber\\ 
 O_4 &= \frac{1}{2}\partial_\mu\phi\partial_\nu\phi {F^{\mu}}_\rho F^{\nu\rho} \,. 
\end{align}
The notation is such that $\bar{\psi}\gamma^\nu\overset{\leftrightarrow}{D^\mu}\psi \equiv \frac{1}{2}\bar{\psi}\gamma^\nu D^\mu\psi- \frac{1}{2}\overline{(D^\mu \psi)}\gamma^\nu \psi$,  $D_\mu\equiv \partial_\mu-i e A_\mu$ with $A_\mu$ the vector potential of the photon and $e$ the charge of electron, and $F_{\mu\nu}=\partial_\mu A_\nu-\partial_\nu A_\mu$ is the gauge invariant EM field tensor. 

The operators $O_1$ and $\overline{O}_1$ can be transformed into each other using the EoM $i\gamma^\mu D_\mu \psi=m_\psi \psi$. This means that by a redefinition of the field $\psi$, one can eliminate $\overline{O}_1$, at the expense of shifting the Wilson coefficient of $O_1$ (see Appendix \ref{appendix:operator elimination} for more details). Due to the soft breaking of shift symmetry from the mass term, there could also be operators of the form  $\phi^2 i \bar{\psi}\gamma^\nu \overset{\leftrightarrow}{D_\nu}\psi$ and $\phi^2 F_{\nu\rho} F^{\nu\rho}$, suppressed by $m_\phi^2/\Lambda^2$. When taking $\phi$ as the DM background, these two operators have the same effects as $O_1$ and $O_3$ for the experiments we consider. They will not mix with the dimension-8 operators under RG evolution. We therefore  ignore these two operators in most of the following discussion, but point out when they are relevant. 

Thus our effective Lagrangian is given by 
\begin{equation}
\label{eq:L_EFT}
    \mathcal{L}_\text{eff}=i\bar{\psi}(\slashed{D}-m_\psi)\psi+\frac{1}{2}\partial_\mu \phi \partial^\mu \phi-\frac{1}{2}m_\phi^2 \phi^2+\sum_{i=1}^4 c_i O_i\,.
\end{equation}
Notice that with this definition the Wilson coefficients
$c_i$ carry mass dimension minus four.  When interpreting
the bounds on the Wilson coefficients from experimental data discussed later on, it will sometimes be useful to make the mass dependence of the Wilson coefficients explicit by writing
\begin{align}
\label{eq:CtoM}
c_i = \frac{16\pi^2}{\Lambda^4} \,,
\end{align}
where $\Lambda$ is a mass scale and the factor of $16\pi^2$
follows the conventions of naive dimensional analysis (see e.g.\cite{Jenkins:2013sda}). Using this equation,  a given bound on a Wilson coefficient can be translated into a mass scale being indirectly probed by the measurement.

\subsection{Renormalization-group flow} \label{sec:RG flow}
The Wilson coefficients $c_i \equiv c_i(\mu)$ are functions of the renormalization scale $\mu$. From the top-down point of view, they are obtained by matching the EFT and the UV theory at the scale $\Lambda$. Then to use them for processes at
a typical low-energy scale $\mu_{\rm exp}$ defined by experiments, one needs to evolve these coefficients from $\mu=\Lambda$ down to $\mu=\mu_{\rm exp}$. This evolution is governed by the Renormalization Group Equations (RGE).  

The computation of the RG flow is outlined in Appendix \ref{appendix:RGflow}. For the RGE at one-loop level we find
\begin{gather}
\label{eq:RGE_CI}
 \renewcommand\arraystretch{1.5}
 \frac{\rm d}{\rm d \log \mu}\begin{bmatrix}
     c_1 \\ c_2 \\ c_3  \\ c_4
 \end{bmatrix}
 = \frac{\alpha}{4\pi}
  \begin{bmatrix}
   0 & \frac{-4}{3} &  12 & \frac{7}{3}  \\
   0 & \frac{16}{3} & 0  & \frac{8}{3}\\
    0 & \frac{-4}{3}  & \frac{8}{3}  & 0 \\
    0 &  \frac{16}{3} &  0 & \frac{8}{3} \\
   \end{bmatrix}
   \begin{bmatrix}
    c_1 \\ c_2 \\ c_3 \\ c_4
   \end{bmatrix}\, ,
\end{gather}
where $\alpha=e^2/(4\pi)$ is the fine structure constant.
The non-zero off-diagonal terms in the above matrix imply that the operators generally mix with each other under RG evolution. However, we notice that non-trivial patterns exist in our case and are governed by symmetries.
This can be seen explicitly by changing the operator basis and organizing the operators into two sectors, characterized by the quantum number $j$ of the angular momentum of the $(\phi,\phi)$ channel~\cite{Jiang:2020rwz}:

\begin{itemize}
    \item $j=0$ sector:
    \begin{align}
        \mathcal{O}_1&\equiv O_1=\left(\partial_\mu\phi\partial^\mu\phi\right) i \bar{\psi}\gamma^\nu \overset{\leftrightarrow}{D_\nu}\psi\,, \nonumber \\
        \mathcal{O}_3&\equiv O_3 =\left( \partial_\mu\phi\partial^\mu\phi\right) \frac{1}{2}F_{\nu\rho} F^{\nu\rho}\,.
        \label{jeq0}
    \end{align}
    \item $j=2$ traceless sector:
    \begin{align}
         \mathcal{O}_2&\equiv O_2-\frac{1}{4}O_1 = \left(\partial_\mu\phi\partial_\nu\phi-\frac{1}{4}\eta_{\mu\nu}\partial_\rho\phi\partial^\rho\phi\right) i \bar{\psi}\gamma^\nu\overset{\leftrightarrow}{D^\mu}\psi\,, \nonumber \\
         \mathcal{O}_4&\equiv O_4-\frac{1}{4}O_3 = \left(\partial_\mu\phi\partial_\nu\phi-\frac{1}{4}\eta_{\mu\nu}\partial_\rho\phi\partial^\rho\phi\right)\frac{1}{2} {F^{\mu}}_\rho F^{\nu\rho}\,. \label{jeq2}
     \end{align}
\end{itemize}
The $j=0$ sector is easy to identify since its operators are products of two scalar operators in the dark and QED sectors, respectively.  The spin-2 sector can be checked by noticing that the contribution from $\mathcal{O}_2$($\mathcal{O}_4$) to the amplitude for  $\phi+\phi\rightarrow \psi+\psi$ ($\phi+\phi\rightarrow \gamma+\gamma$) scattering consists only of partial waves of angular momentum $j=2$. We note that in the $j=2$ operators, the two $\phi$ fields construct a traceless tensor.

Written in the new basis, the mixing matrix is (with $\mathcal{C}_i$ being the coefficient of $\mathcal{O}_i$)
\begin{gather}\label{mixingmatrix}
 \renewcommand\arraystretch{1.5}
 \frac{\rm d}{\rm d \log \mu}\begin{bmatrix}
     \mathcal{C}_1 \\ \mathcal{C}_3 \\ \mathcal{C}_2  \\ \mathcal{C}_4
 \end{bmatrix}
 =  \frac{\alpha}{4\pi }
  \begin{bmatrix}
   \textcolor{blue}{0} & \textcolor{blue}{12} & 0  & 0  \\
   \textcolor{blue}{0} & \textcolor{blue}{\frac{8}{3}} & 0  & 0\\
    0 & 0 & \textcolor{red}{\frac{16}{3}} & \textcolor{red}{\frac{8}{3}} \\
    0 & 0 & \textcolor{red}{\frac{16}{3}} & \textcolor{red}{\frac{8}{3}} \\
   \end{bmatrix}
   \begin{bmatrix}
    \mathcal{C}_1 \\ \mathcal{C}_3 \\ \mathcal{C}_2 \\ \mathcal{C}_4
   \end{bmatrix} \,.
\end{gather}
The mixing matrix is now block diagonal (colour coded in Eq.~(\ref{mixingmatrix}) with blue and red), indicating that RG flow between the two sectors is forbidden by angular momentum conservation. We will work with the new basis defined by $\mathcal{O}_i$ in the following. 

There are two additional noteworthy features of the above mixing matrix. First, in the spin-0 sector (colour coded with blue), the operator $\mathcal{O}_1$ does not generate $\mathcal{O}_3$, whereas $\mathcal{O}_3$ does generate $\mathcal{O}_1$. This is because $\mathcal{O}_1$ is effectively dimension 7 (see Appendix \ref{appendix:operator elimination}) whereas $\mathcal{O}_3$ is dimension 8, and in QFT there is no flow from an operator of lower dimension to an operator of higher dimension. Second, in the $j=2$ sector, the 2$\times$2 mixing matrix (colour coded with red) has rank 1, suggesting that there is a linear combination of operators, namely $\mathcal{O}_2-2\mathcal{O}_4$, that does not run. Whether this holds only at one loop level or is also due to some hidden symmetry remains to be verified. 

In section~\ref{sec:bounds} we will derive bounds on the Wilson
coefficients from terrestrial experiments that 
are sensitive to scales at or below the electron mass.
The Wilson coefficients $\mathcal{C}_i(m_e)$ entering
such bounds are related to those at the scale 
$\Lambda$ characteristic of a UV completion
through the solution to the RG equation.  Neglecting the running of $\alpha$
this solution is simply
\begin{align}
\label{eq:RG_sol}
  \mathcal{C}_i(m_e) & = \left[ \exp\left( \frac{\alpha}{4\pi } \gamma \ln\frac{m_e}{\Lambda}  \right)\right]_{ij}\mathcal{C}_j(\Lambda) \, ,
\end{align}
 where $\gamma$ is the matrix on the right-hand side 
 of Eq.~(\ref{mixingmatrix}).  Taking $\alpha=1/137$ 
 and $\Lambda=1$~TeV as an example, the Wilson
 coefficients at the two scales are related by
\begin{gather}\label{RGsol}
 \renewcommand\arraystretch{1.5}
 \begin{bmatrix}
     \mathcal{C}_1(m_e) \\ \mathcal{C}_3(m_e) \\ \mathcal{C}_2(m_e)  \\ \mathcal{C}_4(m_e)
 \end{bmatrix}
 =  
   \begin{bmatrix}
    \mathcal{C}_{1}(\Lambda) - 0.10 \mathcal{C}_{3}(\Lambda) \\ 0.98 \mathcal{C}_3(\Lambda) \\ 0.96 \mathcal{C}_2(\Lambda) - 0.02\mathcal{C}_4(\Lambda)  \\ 0.98\mathcal{C}_4(\Lambda) - 0.04 \mathcal{C}_2(\Lambda)
   \end{bmatrix} \,. % \qquad \Lambda = \mbox{1~TeV} \,.
\end{gather}
Even for the modest value $\Lambda=1$~TeV, the effects of
RG running are as large as 5--10\%, depending on the coefficient.  

\subsection{An example UV completion}
\label{sec:uvcomp}
The form of the operators, in particular in the $j=2$ sector, indicates that they can originate from integrating out heavy spin-2 particles from a UV theory. Massive spin-2 particles generally arise as composite states (analogous to spin-2 mesons in QCD), 
or in models with extra dimensions such as the Kaluza-Klein (KK) graviton modes~\cite{Kaluza:1921tu,Klein:1926fj}. 
Well known examples are the ADD~\cite{Arkani-Hamed:1998jmv} model and the RS model~\cite{Randall:1999ee}, motivated as solutions to the hierarchy between the Planck scale and the weak scale.  In the ADD setup, the extra dimensions are flat and the fundamental scale is on the order of a TeV. The weakness of gravity is due its leakage into the extra dimensions and to get an effective Planck scale in 4D, the size of extra dimensions are relatively large. So in the ADD case the KK gravitons are light and the mass gap is small, and show up as collective effects. In the RS model, the extra dimension is exponentially warped by the bulk cosmological constant while the 4D branes at the two boundaries remain flat due to the balance between the cosmological constant and the brane tension. With the fundamental scale of the theory set to be Planck scale, fields living on the IR brane (the one at the more curved end of the extra dimension) will appear to have effective mass of the weak scale. Contrary to the ADD case, the KK graviton modes in RS have order TeV mass gaps and could show up as distinct resonances at colliders. Below we are not necessarily committed to any of these specific scenarios, rather we  show that they can generate the dimension-8 operators contained in our low-energy EFT and thus provide examples of UV completions.

\begin{figure}
    \centering 
%     \feynmandiagram [horizontal=a to b] {
%   i1 -- [fermion] a -- [fermion] i2,
%   a -- [graviton, edge label={\tiny KK-graviton}] b,
%   f1 -- [scalar] b -- [scalar] f2,
% }; 
%  \begin{tikzpicture}
% \node at (2,0) {$\rightarrow$};
% \begin{scope}[xshift=4.cm]
% \begin{feynman}
% \node[circle,fill=black,inner sep=2pt] (a);
% \vertex [above left=of a] (i1);
% \vertex [above right=of a] (f1);
% \vertex [below left=of a] (i2);
% \vertex [below right=of a] (f2);
% \diagram* {
% (f2) -- [scalar] (a)  -- [scalar] (f1);
% (i2) -- [fermion] (a) -- [fermion] (i1);
% };
% \end{feynman}
% \end{scope}
% \end{tikzpicture}
\includegraphics[scale=0.8]{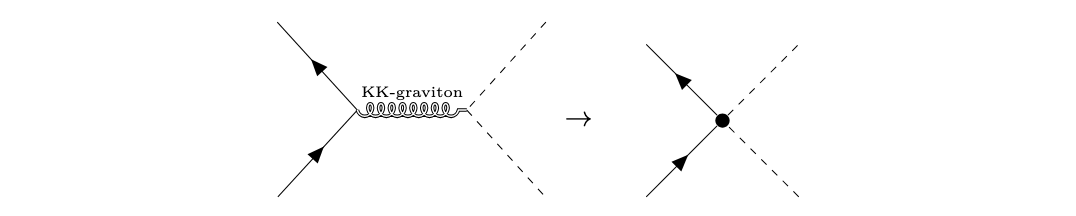}
\caption{At energies much lower than the mass of the KK graviton,  tree-level exchange is approximated by a local interaction and represented as an effective operator.}
\label{fig:EFT_KK}
\end{figure}
We thus consider the tree-level exchange of a massive graviton (e.g. the lightest KK modes in the RS model) between the light scalar $\phi$ and electrons/photons. At energies much smaller than the mass of the KK graviton, this interaction can be approximated by local effective operators, see Fig.~\ref{fig:EFT_KK} as an illustration.

This tree-level matching process can be formally carried out by replacing the heavy field in the action by its classical solution of the EoM. Regardless of its UV origin, the kinetic part of the action of a massive spin-2 field $h^{\mu\nu}$ is described by the Fierz-Pauli action~\cite{Fierz:1939ix}, 
\begin{equation}
\begin{aligned}
S^{\rm FP}=\int d^{4} x\left\{\frac{1}{2} \partial_{\rho} h_{\mu \nu} \partial^{\rho} h^{\mu \nu}\right.&-\partial_{\mu} h_{\nu \rho} \partial^{\nu} h^{\mu \rho}+\partial_{\mu} h^{\mu \nu} \partial_{\nu} h \\
&\left.-\frac{1}{2} \partial_{\rho} h \partial^{\rho} h-\frac{1}{2} m_{KK}^{2}\left(h_{\mu \nu} h^{\mu \nu}-h^{2}\right)\right\}\,,
\end{aligned}
\end{equation}
where $h=h^\mu_\mu$. The KK graviton couples to matter fields through the energy-momentum tensor $T^{\mu\nu}$, with the interaction
\begin{equation}
\mathcal{L}_{\text{int}}=-\frac{1}{\Lambda_{KK}} h_{\mu \nu} T^{\mu \nu}\,.
\end{equation}
Here the scale $\Lambda_{KK}$ characterizes the strength of the interaction. In the RS model we have $\Lambda_{KK} \sim$ TeV but more generally it is associated 
with the scale of spontaneous CFT breaking or the dynamical scale of confinement in composite models. 
For fermions ($\psi$), vectors ($\gamma$),  and scalars ($\phi$), the energy-momentum tensor takes the form
\begin{align}
T_{\mu \nu}^{\psi}&=\frac{i}{4} \bar{\psi}\left(\gamma_{\mu} D_{\nu}+\gamma_{\nu} D_{\mu}\right) \psi-\frac{i}{4}\left(D_{\mu} \bar{\psi} \gamma_{\nu}+D_{\nu} \bar{\psi} \gamma_{\mu}\right) \psi\nonumber\\
&-\eta_{\mu \nu}\left(\bar{\psi} \gamma^{\rho} D_{\rho} \psi-m_{\psi} \bar{\psi} \psi\right)+\frac{i}{2} \eta_{\mu \nu} \partial^{\rho}\left(\bar{\psi} \gamma_{\rho} \psi\right)\,,\\
T_{\mu \nu}^{\gamma}&=\frac{1}{4} \eta_{\mu \nu} F^{\lambda \rho} F_{\lambda \rho}-F_{\mu \lambda} F_{\nu}^{\,\lambda}\,,\\
T_{\mu \nu}^{\phi}&=\partial^{\mu} \phi \partial^{\nu} \phi-\frac{1}{2} \eta^{\mu \nu}\left(\partial_{\rho} \phi \partial^{\rho} \phi-m_{\phi}^{2} \phi^{2}\right)\,.
\label{eq:Tmunu}
\end{align}

To integrate out the KK graviton at tree level, we rewrite the full action as 
\begin{equation}
S=\int d^{4} x \left\{\frac{1}{2} h_{\mu \nu} \mathcal{O}^{\mu \nu \rho \sigma} h_{\rho \sigma}-\frac{1}{\Lambda_{KK}} h_{\mu \nu} T^{\mu \nu} \right\}\,,
\end{equation}
with 
\begin{equation}
\mathcal{O}_{\rho \sigma}^{\mu \nu}=-\left[\eta_{(\rho}^{(\mu} \eta_{\sigma)}^{\nu)}-\eta^{\mu \nu} \eta_{\rho \sigma}\right]\left(\square+m_{KK}^{2}\right)+2 \partial^{(\mu} \partial_{(\rho} \eta_{\sigma)}^{\nu)}-\partial^{\mu} \partial^{\nu} \eta_{\rho \sigma}-\partial_{\rho} \partial_{\sigma} \eta^{\mu \nu}\,.
\end{equation}
 Then the classical equation of motion for $h_{\mu\nu}$ reads
\begin{equation}
\mathcal{O}^{\mu \nu \rho \sigma} h_{\rho \sigma}=\frac{1}{\Lambda_{KK}}  T^{\mu \nu}\,,
\end{equation}
and we obtain the classical solution 
\begin{equation}
 h^c_{\rho \sigma}=\mathcal{O}^{-1}_{ \rho \sigma \alpha \beta}\frac{1}{\Lambda_{KK}}  T^{\alpha \beta}\,,
\end{equation}
with $\mathcal{O}^{-1}$ defined by
\begin{equation}
\mathcal{O}^{\mu \nu \rho \sigma} \mathcal{O}^{-1}_{ \rho \sigma \alpha \beta}=\frac{1}{2}  (\delta^\mu_\alpha \delta^\nu_\beta +\delta^\mu_\beta \delta^\nu_\alpha)\,.
\end{equation}
The effective Lagrangian is then obtained by substituting the classical solution back into the original Lagrangian,
\begin{align}
\mathcal{L}^{\text{EFT}}&=-\frac{1}{2\Lambda_{KK}} h^c_{\mu \nu} T^{\mu \nu}\nonumber\\
&=-\frac{1}{2\Lambda_{KK}^2} T^{\mu \nu}\mathcal{O}^{-1}_{\mu \nu \rho \sigma } T^{\rho \sigma}\,.
\label{eq:effL}
\end{align}
The effective operators are obtained by expanding $\mathcal{O}^{-1}$ in $1/m_{KK}^2$. To leading order, we have 
 \begin{align}
 \mathcal{O}^{-1}_{\mu \nu \rho \sigma }=-\frac{1}{m_{KK}^2}\left(\frac{1}{2}\eta_{\mu \rho} \eta_{\nu \sigma}+\frac{1}{2}\eta_{\mu \sigma} \eta_{\nu \rho}-\frac{1}{3} \eta_{\mu \nu} \eta_{\rho \sigma}\right)\,.
\end{align}

Substituting the expressions for $T^{\mu\nu}$ in Eq.~\eqref{eq:Tmunu} into Eq.~\eqref{eq:effL}, we obtain the effective operators
\begin{align} \label{KK_EFT}
\mathcal{L}^{\text{EFT}}&\supset \frac{1}{2\Lambda_{KK}^2m_{KK}^2}[(\partial^{\mu} \phi \partial^{\nu} \phi-\frac{1}{6} \eta^{\mu \nu}(\partial_{\rho} \phi \partial^{\rho} \phi+m_{\phi}^{2} \phi^{2}))i \bar{\psi}\gamma^\nu \overset{\leftrightarrow}{D_\mu}\psi \nonumber\\
&-(\partial^{\mu} \phi \partial^{\nu} \phi-\frac{1}{4} \eta^{\mu \nu}\partial_{\rho} \phi \partial^{\rho} \phi)F_{\mu \lambda}F^{\nu \lambda}]\,.
\end{align}
In the above calculation, we have ignored terms in $T^{\psi}_{\mu\nu}$ that vanish under the EoM. Matching with the effective Lagrangian, we find
\begin{equation}
\label{eq:KK_C_res}
    \mathcal{C}_1=\frac{1}{24\Lambda_{KK}^2m_{KK}^2}\,,\quad \mathcal{C}_2=\frac{1}{2\Lambda_{KK}^2m_{KK}^2}\,,\quad \mathcal{C}_3=0\,,\quad \mathcal{C}_4=-\frac{1}{2\Lambda_{KK}^2m_{KK}^2}\,.
\end{equation}

We see that in this specific example, the UV scale is the
hybrid scale $\sqrt{\Lambda_{KK}m_{KK}}$, which could be the geometrical mean of the extra-dimension's volume and fundamental scale in flat extra-dimension models or just the scale of CFT breaking or confinement in RS and composite models respectively.  

As a final comment, we note that in models of spontaneously broken CFTs a similar coupling of the form of $\partial^{\mu} \phi \partial^{\nu} \phi \times T_{\mu\nu}/f_{\rm CFT}^4$, where $\phi$ is the dilaton field, $f_{\rm CFT}$ being the scale of spontaneous CFT breaking, is predicted to be present with the coefficient of this term predicted by the $a$-anomaly~\cite{Csaki:2022htl,Komargodski:2011vj}.

\subsection{Positivity bounds}

The Wilson coefficients of the effective operators are constrained by fundamental properties of the UV theory, such as Lorentz invariance, unitarity, causality and locality. These properties ensure that the S-matrix is an analytic function of the Mandelstam variables $(s,t,u)$, other than poles and cuts determined by on-shell unitarity.    The simplest bounds are obtained by considering the two-to-two scattering amplitude $\mathcal{A}(s)$ in the forward limit in the complex plane of the $s$ (see e.g.~\cite{Adams:2006sv, Remmen:2019cyz}). The Wilson coefficient is related to a contour integral of the amplitude around the origin, $\oint \frac{d s}{2 \pi i} \frac{\mathcal{A}(s)}{s^3}$. By distorting the contour, this integral becomes the integral of the discontinuity along the real axis up to infinity and unitarity, which relates the discontinuity to the cross section, ensures its positivity. In the end, this positivity bound is expressed by
\begin{equation}
    \frac{\rm d^2 \mathcal{A}(s)}{\rm d s^2}\geq 0.
\end{equation}

To apply this to our case, we consider the contribution of the effective operators to the forward scattering of $\phi+\psi \rightarrow \phi+\psi$ and $\phi+\gamma \rightarrow \phi+\gamma$, in which case we obtain
\begin{equation}
    \frac{\rm d^2 \mathcal{A}(\phi+\psi \rightarrow \phi+\psi)}{\rm d s^2}=2\mathcal{C}_2 \,,\quad  \frac{\rm d^2 \mathcal{A}(\phi+\gamma \rightarrow \phi+\gamma)}{\rm d s^2}=-\mathcal{C}_4/2.
\end{equation}
 The positivity bounds thus require
 \begin{equation}
 \label{eq:posi_bounds}
    \mathcal{C}_2\geq 0\,, \qquad  \mathcal{C}_4 \leq 0.
\end{equation}   
Notice that the  operators $\mathcal{O}_1$ and $\mathcal{O}_3$ do not have contributions of order $s^2$ to the scattering amplitude and are 
thus not constrained by the positivity bound. The bounds above are consistent with the Wilson coefficients obtained from integrating out the KK graviton, see Eq.~(\ref{eq:KK_C_res}).

\section{Experimental bounds on the operators} 
\label{sec:bounds}
In this section we study the experimental implications of our EFT, assuming that $\phi$ is a viable ultralight, sub-eV, dark matter candidate. In the rest frame of the solar system, the field $\phi$ can be approximated as a background field,
\begin{equation}
\label{eq:wave}
\phi(t,\vec{x})=\phi_0\text{ cos}(m_\phi(t-\vec{v}_{\rm DM}\cdot\vec{x}))\,,
\end{equation}
where $|\vec{v}_{\rm DM}|\simeq10^{-3}$~\cite{Sch_nrich_2010,Eilers_2019} is the relative velocity of the DM wind to the sun. The amplitude $\phi_0$ can be determined by noticing that the energy density of the $\phi$ field is \begin{align}
\rho_\phi=\frac{1}{2}(\dot{\phi}^2+m_\phi^2\phi^2)= \frac{1}{2} m_{\phi}^2 \phi_0^2 \, .
\end{align} 
Identifying this with the local DM density, $\rho_\phi = \rho_{\rm DM}\simeq0.4$~GeV/cm$^3$ \cite{Catena2010}, we have $\phi_0=\sqrt{2\rho_{\rm DM}}/m_{\phi}$. Notice that for simplicity we neglect here the stochastic nature of the DM field~\cite{Centers:2019dyn,Lisanti:2021vij,Foster:2017hbq}, as well as the DM velocity dispersion, as these effects lead to subdominant effects below.  

In cases where the spatial dependence of Eq.~(\ref{eq:wave}) can be neglected, the effective operators induce time oscillations in fundamental constants such as the speed of light, the electron mass, and the fine structure constant. 
These effects are the subject of Section~\ref{sec:fund_oscillations}.  In 
experiments testing for  directional Lorentz violation, on 
the other hand, the spatial dependence is crucial and the $j=2$ operators lead to the effects discussed in Section~\ref{sec:bounds_C24}.

Current experiments probing both kinds of effects have produced null results and can be used to set bounds on the Wilson coefficients as described below.  The characteristic energy scales of such experiments are at or below the electron mass, so the bounds can be interpreted as being on $\mathcal{C}_i(m_e)$ -- they can be turned into bounds on Wilson coefficients evaluated at high-energy scales of potential UV completions using Eq.~(\ref{eq:RG_sol}).

\subsection{Bounds from the oscillation of fundamental constants}
\label{sec:fund_oscillations}
We start by noticing that, by replacing $\phi$ with its background 
value in Eq.~(\ref{eq:wave}), $\mathcal{O}_1$ and $\mathcal{O}_3$ lead to corrections to the electron mass term and photon kinetic term respectively~\footnote{The shift symmetry breaking operators $\mathcal{Q}_1=m_\phi^2\phi^2 i \bar{\psi}\gamma^\nu \overset{\leftrightarrow}{D_\nu}\psi$ and $\mathcal{Q}_3=\frac{1}{2}m_\phi^2\phi^2 F_{\nu\rho} F^{\nu\rho}$, with coefficients $d_1$ and $d_2$, 
can be taken into account by interpreting the bounds derived for $\mathcal{C}_1$ and $\mathcal{C}_3$ as being on the combinations of them.}
\begin{align}
\label{eq:O1&O3}
    \mathcal{C}_1 \mathcal{O}_1 &\longrightarrow \mathcal{C}_1\rho_{\rm DM}(1-\cos(2 m_\phi t))m_e \bar\psi \psi \, ,\\
    \mathcal{C}_3 \mathcal{O}_3 &\longrightarrow \frac{1}{2} \mathcal{C}_3 \rho_{\rm DM} (1-\cos(2m_\phi t))  F_{\mu\nu}F^{\mu\nu}.
       \end{align}
In deriving these equations we have ignored the spatial dependence inside the cosine, 
since the wavelength of the $\phi$ field is always much larger than the typical size of the experiments we consider in the following, and also used the EoM, $i\slashed{D}\psi=m_e \psi$. 
We find that $\mathcal{O}_1$ induces a shift in the electron mass, while $\mathcal{O}_3$ induces a shift of the fine structure constant (after a field redefinition to bring the photon kinetic term into the canonical form), both oscillate with frequency $2m_\phi$:
\begin{align}
\label{eq:me&alpha}
    \frac{\Delta m_e}{m_e}&= -\mathcal{C}_1\rho_{\rm DM}(1-\cos(2 m_\phi t)) \, ,\\
    \frac{\Delta \alpha}{\alpha} &=2 \mathcal{C}_3 \rho_{\rm DM}(1-\cos(2 m_\phi t)) \, .
\end{align}

Next we consider the effects of $\mathcal{O}_2$ and $\mathcal{O}_4$. With the DM background in Eq.~(\ref{eq:wave}) (ignoring again the spatial dependence), 
we have
\begin{align}
\label{eq:O2&O4}
    \mathcal{C}_2 \mathcal{O}_2 &\longrightarrow  i C_{\mu \nu} \bar{\psi}\gamma^\mu\overset{\leftrightarrow}{D^\nu}\psi \, ,\\
    \mathcal{C}_4 \mathcal{O}_4 &\longrightarrow -\frac{1}{2} K_{\mu \nu}  F^{\mu\rho}{F^{\nu}}_{\rho},
\end{align}
with 
\begin{align}
C_{\mu \nu} & =\mathcal{C}_2\rho_{\rm DM}\sin^2(m_\phi t) \times {\rm Diag} \{3/2,1/2,1/2,1/2\} \, , \nonumber \\
K_{\mu \nu} & =-\mathcal{C}_4\rho_{\rm DM}\sin^2(m_\phi t) \times {\rm Diag} \{3/2,1/2,1/2,1/2\} \, .
\end{align}
We see that $\mathcal{O}_2$ and $\mathcal{O}_4$ modify the kinetic terms of the electron and photon fields in a way that discriminates between time and spatial components, and thus their effects can not simply be taken as a shift to fundamental constants. 

In low-energy atomic physics experiments the operator $\mathcal{O}_2$ induces non-trivial changes to the non-relativistic QED Hamiltonian, $h$.  For instance, the kinetic term for the electron is shifted to
\begin{align}
 h\supset -\left(1-\frac{5}{2}\mathcal{C}_2 \rho_{\rm DM} \sin^2(m_\phi t)\right)\frac{\nabla^2}{2m_e}.
\end{align}
This leads to a change in the Rydberg energy, which can be obtained by 
replacing $m_e \to (1+\frac{5}{2}\mathcal{C}_2\rho_{\rm DM}\sin^2(m_\phi t)) m_e$. However, this simple replacement is not generally true for higher-order terms responsible for (hyper)fine structure, such as spin-orbit and spin-spin interactions. We discuss the shifts in these terms in Appendix~\ref{sec:H_levels}.

To see the effect of $\mathcal{O}_4$, we write the photon Lagrangian in the following form:
\begin{align}
\label{eq:L_photon_mod}
     \mathcal{L}=&-\frac{1}{4}F_{\mu\nu}F^{\mu\nu}+\mathcal{C}_4 \mathcal{O}_4 \\
     \to &\frac{1}{2}(1-\mathcal{C}_4\rho_{\rm DM} \sin^2(m_\phi t))E^2 - \frac{1}{2}(1+\mathcal{C}_4\rho_{\rm DM} \sin^2(m_\phi t))B^2 +\dots \,,
\end{align}
where the $\dots$ refers to terms suppressed by $\vec{v}_{\rm DM}$, which are the topic of section~\ref{sec:bounds_C24}. 
We see that the DM background acts as a medium with effective permittivity $\epsilon$ and permeability $\mu$ given 
by 
\begin{align}
\label{eq:eps_and_mu}
\epsilon= \frac{1}{\mu} = 1-\mathcal{C}_4\rho_{\rm DM}\sin^2(m_\phi t)\, .
\end{align}

This implies  that the speed of light is modified to
\begin{align}
\label{eq:c_mod}
    c'&=\frac{1}{\sqrt{\epsilon \mu}}  \simeq 1+\mathcal{C}_4 \rho_{\rm DM} \sin^2(m_\phi t).
\end{align}
The requirement of $c'\leq1$ gives $\mathcal{C}_4 \leq 0$, in agreement with the positivity bound in Eq.~(\ref{eq:posi_bounds}).  The coefficient $\mathcal{C}_4$ also affects atomic-physics observables such as the Rydberg energy and (hyper)fine structure -- the relevant calculations are given in  Appendix~\ref{sec:H_levels}.

\begin{figure}
     \centering
\includegraphics[width=.45\textwidth]{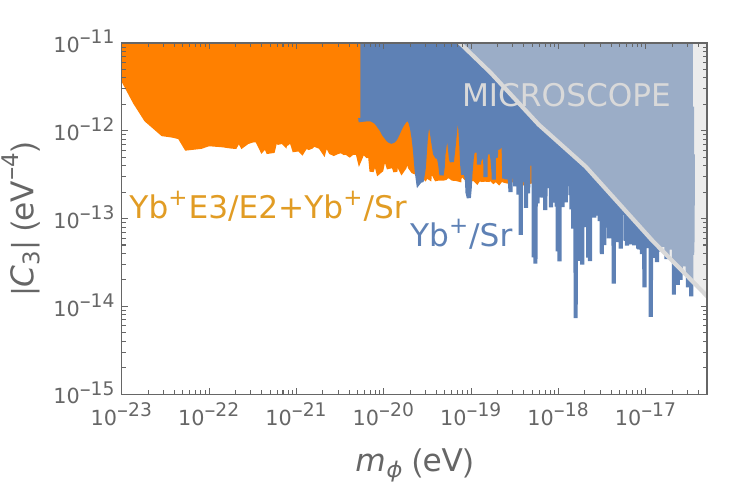}~
\includegraphics[width=.45\textwidth]{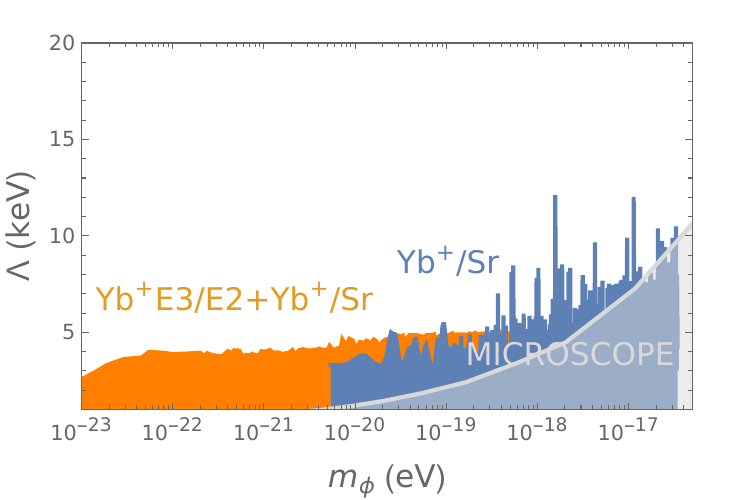}~\\
\includegraphics[width=.45\textwidth]{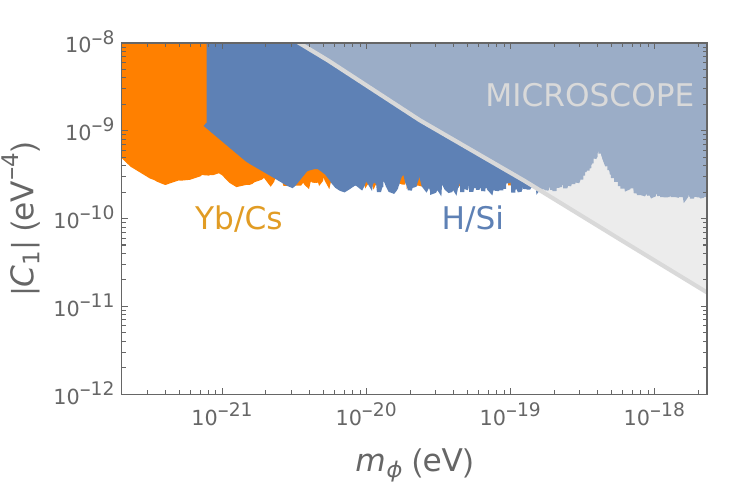}~
\includegraphics[width=.45\textwidth]{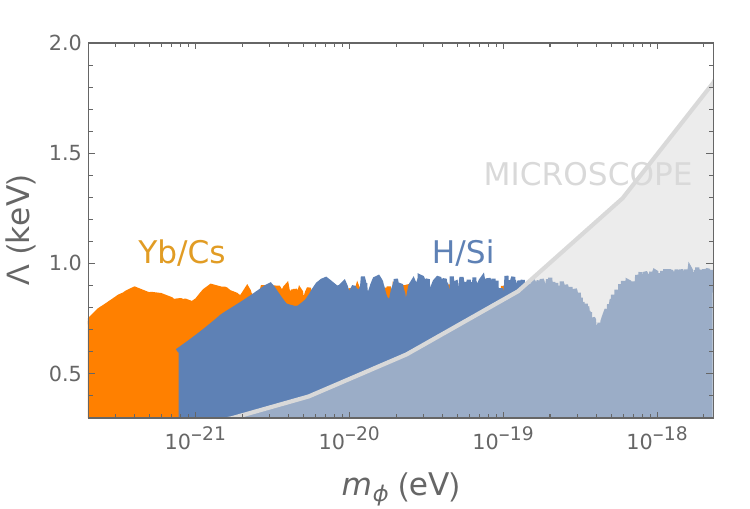}
     \caption{The bounds on Wilson coefficients from experiments searching for oscillations of fundamental constants for different dark matter mass $m_\phi$. Upper: Exclusion regions of $\mathcal{C}_3$ (left) and the corresponding cut-off $\Lambda\sim \sqrt{4\pi}\mathcal{C}_3^{-1/4}$ (right). Orange regions are obtained from \cite{Filzinger:2023zrs} (comparison of Yb$^+$ E3/E2 as well as Yb$^+$/Sr) and blue regions from \cite{Sherrill:2023zah} (comparison of Yb$^+$/Sr). Lower: Exclusion regions of $\mathcal{C}_1$ (left) and the corresponding cut-off $\Lambda\sim \sqrt{4\pi} \mathcal{C}_1^{-1/4}$ (right), obtained by comparing the frequencies of Yb/Cs~\cite{Kobayashi:2022vsf} (orange) or H/Si~\cite{Kennedy2020} (blue). Also shown with gray shaded regions are bounds from the MICROSCOPE experiment, obtained from direct translation of the bounds given in \cite{Banerjee:2022sqg, Hees:2018fpg}.}
     \label{fig:clockbounds}
 \end{figure}

We now consider experimental probes of the effects discussed above. 
Oscillations of the electron mass and fine structure constant due to $\mathcal{O}_1$ and $\mathcal{O}_3$ can be probed by comparing two systems which have typical frequencies $f_1$ and $f_2$ with different dependence on $m_e$ and $\alpha$, and looking for the sinusoidal modulations of the ratio of the two frequencies. Currently the strongest bounds on the oscillation of $\alpha$ come from optical clock comparisons. In \cite{Lange:2020cul,Filzinger:2023zrs}, the $^2S_{1/2}(F = 0) \rightarrow$       $^2D_{3/2}(F = 2)$ electric quadrupole (E2) and the $^2S_{1/2}(F = 0) \rightarrow$  $^2F_{7/2}(F = 3)$ electric octopole (E3) transitions of $^{171}$Yb$^+$ are measured and compared. The strong bounds benefit from the strong $\alpha$ dependence $\frac{\Delta (f_{E3}/f_{E2})}{f_{E3}/f_{E2}}=-6.95\frac{\Delta \alpha}{\alpha}$. Also studied in \cite{Filzinger:2023zrs}, as well as in \cite{Sherrill:2023zah}, is the comparison between the E3 transition with the $^1S_{0} \rightarrow$ $^3P_{0}$ transition in $^{87}$Sr, which has the sensitivity  $\frac{\Delta (f_{E3}/f_{Sr})}{f_{E3}/f_{Sr}}=-6.01\frac{\Delta \alpha}{\alpha}$.  Considering the scenario where the oscillation of $\alpha$ is due to a dark matter background $\varphi$, with interaction of the form $\mathcal{L}\supset d_{e}\sqrt{4\pi G}\varphi F_{\mu\nu}F^{\mu\nu}/4$ where $G$ is the Newton's constant, bounds are reported on the parameter $d_e$ in \cite{Filzinger:2023zrs,Sherrill:2023zah}. By identifying the same amplitude of oscillation of $\alpha$ at the same frequency, we can relate the bound on our Wilson coefficient at given $m_\phi$ to the bound on $d_e$ at $2m_\phi$, with the correspondence $\mathcal{C}_3(m_\phi)\leftrightarrow \sqrt{\frac{2\pi G}{\rho_{\rm DM}}}\frac{d_{e}(2m_\phi)}{2m_\phi}$. We show the bounds on $\mathcal{C}_3$ from these two experiments in the upper left plot in Fig.~\ref{fig:clockbounds}.  We see that $\mathcal{C}_3$ is more constrained by the Yb$^+$/Sr comparison at $m_\phi \gtrsim 10^{-18}$ eV, which requires  roughly
\begin{equation}
      |\mathcal{C}_3|\lesssim  10^{-14} \,{\rm eV}^{-4}.
\end{equation}
 In the upper right plot of Fig.~\ref{fig:clockbounds} we show the corresponding bound on $\Lambda$, using Eq.~(\ref{eq:CtoM}). The bound is very weak, $\Lambda\gtrsim 10$ keV, compared to the scale the UV completions considered in Sec.~\ref{sec:uvcomp}. 

The bounds on the oscillation of $m_e$ are generally weaker than those on $\alpha$. In~\cite{Kennedy2020}, the resonant frequency $f_c$ of a silicon optical cavity and the frequency $f_H$ of the $^1$H hyperfine transition are compared, yielding sensitivity to oscillations of $m_e$ according to $\frac{\Delta (f_{H}/f_{c})}{f_{H}/f_{c}}=\frac{\Delta m_e}{m_e}$. The bound is reported in terms of the parameter $d_{me}$, defined by the interaction $\mathcal{L}\supset -d_{me}\sqrt{4\pi G}m_e \varphi \bar \psi \psi$. At lower masses $m_\phi\lesssim 10^{-20}$ eV, stronger bounds are provided in~\cite{Kobayashi:2022vsf} by the comparison of $^{171}$Yb optical lattice clock and a $^{133}$Cs
fountain clock. Using the correspondence $\mathcal{C}_1(m_\phi)\leftrightarrow\sqrt{\frac{2 \pi G}{\rho_{\rm DM}}}\frac{d_{m_e}(2m_\phi)}{m_\phi}$, we can translate the bounds therein to  bounds on $\mathcal{C}_1$. As shown in the lower left plot of Fig.~\ref{fig:clockbounds}, we find roughly
\begin{equation}
      |\mathcal{C}_1| \lesssim  10^{-10} \,{\rm eV}^{-4} \, .
    \end{equation}
The corresponding bound on the cut-off $\Lambda=\sqrt{4\pi} |\mathcal{C}_1|^{-1/4}$ is shown in the lower right plot, leading to $\Lambda \gtrsim $ keV.

There is one subtlety in re-interpreting bounds derived from linear DM coupling as those on quadratic coupling, as we did above. When the quadratic coupling is positive and sufficiently strong, the DM profile is modified significantly due to 
interactions with the earth and the $\phi$ field gets screened at the surface of the earth \cite{Banerjee:2022sqg,Hees:2018fpg}. Any terrestrial experiments will thus not feel it. For our case, this happens at $\mathcal{C}_{1,3} \gtrsim 10^{-3} (\frac{10^{-20} {\rm eV}}{m_\phi})^2 {\rm eV}^{-4}$, so is not relevant for the parameter space studied above, where the clock experiments are most sensitive. However, at larger $m_\phi$ the screening effect could be important. 

In Fig.~\ref{fig:clockbounds} we also show the bounds from
equivalence principle tests. The stringent limit is from the MICROSCOPE experiment~\cite{Berge:2017ovy} which measures the
differential accelerations of macroscopic masses. We notice that for $\mathcal{C}_3$ this bound is weaker than those from clocks, while for $\mathcal{C}_1$ it could be stronger at $m_\phi\gtrsim 10^{-19}$ eV. However, the bounds from clocks are expected to be improved significantly by future experiments. In particular, the nuclear clock \cite{Flambaum:2006ak,Peik:2020cwm} is expected to improve the ability to probe scalar dark matter by 5-6 orders of magnitude, due to its strongly enhanced sensitivity. 
 
Although the operators $\mathcal{O}_2$ and $\mathcal{O}_4$ lead to more complicated effects than the simple oscillation of fundamental constants, they can still affect the results of aforementioned frequency comparison experiments. Consider for example the $f_c/f_H$ comparison. The frequency of the cavity is determined by $f_c=N c/2 d$, where $N$ is an integer number, $c$ is the speed of light and modified by $\mathcal{C}_4$ according Eq.~(\ref{eq:c_mod}), $d$ is the cavity size and proportional to the Bohr radius $d\propto 1/m_e e^2$ which is modified by $m_e\rightarrow(1+\frac{5}{2}\mathcal{C}_2\rho_{\rm DM}\sin^2(m_\phi t)) m_e$ and $e^2\rightarrow(1+\mathcal{C}_4\rho_{\rm DM}\sin^2(m_\phi t)) e^2$ (from the correction to the Coulomb potential due to the change in permittivity). We find that $f_c \rightarrow  (1+(\frac{5}{2}\mathcal{C}_2+2\mathcal{C}_4)\rho_{\rm DM}\sin^2(m_\phi t)) f_c$. We also derive the frequency of Hydrogen hyperfine splitting from spin-spin interaction in Appendix~\ref{sec:H_levels} and obtain $f_H \rightarrow (1+(5\mathcal{C}_2+2\mathcal{C}_4)\rho_{\rm DM}\sin^2(m_\phi t)) f_H$. In the end we have $\frac{\Delta (f_{H}/f_{c})}{f_{H}/f_{c}}=\frac{5}{4}\mathcal{C}_2 \rho_{\rm DM}(1-\cos(2m_\phi t))$. The bound obtained for $\mathcal{C}_1$ from this experiment should thus be taken as the bound for $\mathcal{C}_1-5\mathcal{C}_2/4$. In principle, one could also derive similar bounds from other experiments mentioned above. However, the dependence of energy levels on $\mathcal{C}_2$ and $\mathcal{C}_4$ in high $Z$ atoms such as $^{133}$Cs is not as simple as for
the Hydrogen atom and requires numerical calculations taking into account the many-body effects and higher order relativistic corrections, which are beyond the scope of this work. 

Concerning the correction to the speed of light in Eq.~(\ref{eq:c_mod}), we may directly put bounds on $\mathcal{C}_4$ from experiments measuring the propagation of light, for instance, the  Lunar Laser Ranging experiment, 
where the time of travel for a round-way trip of the laser between the earth and the moon was measured.  The oscillation in $c$ maps onto
an oscillation in the time of travel, which would be interpreted as a correction to the distance $r$ between earth and the moon, $\Delta r\approx -r \mathcal{C}_4 \rho_{\rm DM} \text{ sin}^2(m_\phi t)$. However, one should bear in mind that the frequency of the clock used in the experiment would also be modified by $\mathcal{C}_4$. Nevertheless, supposing that there is no non-trivial cancellation between the two effects, we can make an estimation of the bound. In~\cite{Zhang:2020wpj}, the bound of the amplitude $A$ for the oscillation of $\Delta r$ with period of one sidereal month was obtained to be $A\lesssim 1$~mm. This gives a bound $|\mathcal{C}_4|\lesssim 10^{-3}$ eV$^{-4}$ at $m_\phi\approx 10^{-21}$ eV. Translating this to a mass scale using
Eq.~(\ref{eq:CtoM}), we have $\Lambda \sim \sqrt{4\pi} |\mathcal{C}_4|^{-1/4}\lesssim 20$ eV, which is very weak compared to the bounds
from clock/cavity comparison experiments. In Sec.~\ref{sec:bounds_C24} we will show that linear corrections in $\vec{v}_{\rm DM}$ which were neglected in Eq.~(\ref{eq:c_mod}) do not  affect round-way trip measurements and this bound still holds. On the other hand, quadratic corrections in $\vec{v}_{\rm DM}$ do not cancel out in a round way trip, and stronger bounds can be obtained using interferometers.

\subsection{Bounds from experiments measuring Lorentz  violation}\label{sec:bounds_C24}

One feature of particular interest in our EFT is the Lorentz violating effects (in particular the asymmetry under spatial rotations) from $\mathcal{O}_2$ and $\mathcal{O}_4$ when $\phi$ takes the background value Eq.~(\ref{eq:wave}) and $\vec{v}_{\rm DM}$ is not ignored. Our EFT thus provides a concrete model to be tested in the large number of experiments testing Lorentz violation in recent years.

Motivated by potential Lorentz violation at the Planck scale, the general low-energy effective theory describing the effects of spontaneous Lorentz symmetry breaking has been developed as the Standard-Model Extension (SME) \cite{Colladay:1996iz,Colladay:1998fq} and has been widely used to interpret results from experiments hunting for Lorentz violation. For  comparison with our EFT, the relevant terms in the SME Lagrangian take the form 
\begin{equation}
\label{eq:SME}
\mathcal{L}_{\rm SME}\supset i(\eta_{\mu \nu }+c_{\mu \nu}) \bar{\psi}\gamma^\mu\overset{\leftrightarrow}{D^\nu}\psi -\frac{1}{4} F_{\mu\nu} F^{\mu \nu}-\frac{1}{4}\left(k_F\right)_{\kappa \lambda \mu \nu} F^{\kappa \lambda} F^{\mu \nu}.
\end{equation}
Here the parameters $c_{\mu\nu}$ and $\left(k_F\right)_{\kappa \lambda \mu \nu}$ characterize Lorentz-violating, in the electron and photon sector respectively. They are defined in the Sun-centred celestial equatorial frame (SCCEF) \footnote{This frame is defined as follows. The $Z$-axis is parallel to the rotational axis of the Earth and directed north. The $X$-axis points from the Sun towards the vernal equinox. The $Y$-axis completes a right-handed system.} and are assumed to be constant. Both of them are defined to be (double) traceless, $c_{\;\mu}^\mu=0$, $\left(k_F\right)_{\quad\mu \nu}^{\mu \nu}=0$. Furthermore, $\left(k_F\right)_{\kappa \lambda \mu \nu}$ has the symmetries of the Riemann tensor.

Ten of the 19 independent coefficients in $\left(k_F\right)_{\kappa \lambda \mu \nu}$  generate a birefringence effect, which is strongly constrained by polarization measurements of photons traveled over cosmological distances. If these ten components are zero, $\left(k_F\right)_{\kappa \lambda \mu \nu}$ can be formulated as
\begin{equation}
\left(k_F\right)_{\kappa \lambda \mu \nu}=\frac{1}{2}(\eta_{\kappa \mu }k_{\lambda \nu}-\eta_{\kappa \nu }k_{\lambda \mu}-\eta_{\lambda \mu }k_{\kappa \nu}+\eta_{\lambda \nu }k_{\kappa  \mu}),
\end{equation}
with $k_{\mu\nu}\equiv\left(k_F\right)_\alpha^{\; \mu \alpha \nu}$, which is symmetric and traceless.

Comparing with our effective operators after substituting $\phi$  with Eq.~(\ref{eq:wave}) yields 
\begin{align}
\label{eq:cmunu&kmunu}
c_{\mu\nu}&=\mathcal{C}_2\rho_{\text{DM}}(v_\mu v_\nu-\eta_{\mu\nu}/4),\nonumber\\
k_{\mu \nu }&=-\mathcal{C}_4\rho_{\text{DM}}(v_\mu v_\nu-\eta_{\mu\nu}/4),
\end{align} 
with $v_\mu=\left(\sqrt{1+v_{\rm DM}^2},\vec{v}_{\rm DM}\right)$. Unlike in the previous subsection, here (to be compared with the bounds for SME parameters) we ignored the oscillation and took the average $\langle\sin^2(m_\phi t)\rangle=1/2$. This is valid in two regimes: either when the oscillation period is much shorter than the typical time scale of the experiment so that the oscillation is averaged out, or when the oscillation is so slow that the amplitude does not change much during the experiment. The intermediate region requires a dedicated analysis of the original experimental data and we leave it for future work. In this subsection we are more focused on the Galilean violating effects, occurring at order $|\vec{v}_{\rm DM}|^2\sim10^{-6}\,.$  

One of the most precise terrestrial tests of Lorentz violation comes from the comparison of the E3 transitions of two $^{171}$Yb$^+$ clocks~\cite{Sanner2019}. These two clocks are placed in magnetic fields with different orientations. The magnetic fields define the quantization axis for these two ions and break the $m_F$ degeneracy in the $^2F_{7/2}(F = 3)$ states. For both ions the $^2F_{7/2}(F = 3,m_F=0)$ states are selected. If no Lorentz violation exists, the frequencies of the transitions in these two clocks should be the same.  With Lorentz violation, on the other hand, the energy levels of the clocks are modified to be different.  First, the $c_{\mu\nu}$ terms directly add to the non-relativistic Hamiltonian of the electron a spherically asymmetric contribution~\cite{Kostelecky:1999mr}. Second, Lorentz violation in the photon sector modifies the electromagnetic potential sourced by the nucleus~\cite{Bailey:2004na}. These two effects are described by the effective Hamiltonian \cite{Kostelecky:1999mr,Hohensee:2013cya}
\begin{equation}
\label{eq:H_LV}
    \delta h=-C_0^{(2)}\frac{p^2-3p_z^2}{6m_e},
\end{equation}
with $C_0^{(2)}=c'_{xx}+c'_{yy}-2c'_{zz}$ and $c'_{\mu\nu}=c_{\mu\nu}-\frac{1}{2}k_{\mu\nu}$.\footnote{The dependence on the particular combination $c^{\mu\nu}-\frac{1}{2}k^{\mu \nu}$ can be understood by the observation that by doing a coordinate transformation $x^\mu \rightarrow x^\mu-\frac{1}{2}k^\mu_{\;\nu} x^\nu$ ~\cite{Bailey:2004na}, the Lorentz-violating terms in the photon sector can be eliminated and the fermion sector changes as $c^{\mu\nu} \rightarrow c^{\mu\nu}-\frac{1}{2}k^{\mu\nu}$~\cite{Altschul:2006zz}. Such a transformation corresponds to a choice of ``rods and clocks" defined by the freely-propagating electromagnetic field in the experiment.} 

Notice that the $c'_{\mu\nu}$ in Eq.~(\ref{eq:H_LV}) is the value in the clock frame while the parameters in Eq.~(\ref{eq:SME}) are defined in the SCCEF frame, so that one has to do a time-dependent Lorentz transformation to relate them due to the orbiting of the earth around the sun as well as the rotation of the earth. The experiment thus has sensitivities not only to the spatial components  $c'_{XY}$ but also to $c'_{TX}$ and $c'_{TT}$ \footnote{Here and in the following we use the uppercase indices for the SCCEF coordinates and lowercase indices for coordinates in the lab frame.}. However, these are suppressed by a small boost factor, $\beta\sim 10^{-4}$ and therefore result in weaker bounds.

With the data from a six-month measurement period, \cite{Sanner2019}  reports bounds on Lorentz-violating parameters. The strongest bounds are put on the spatial components $c'_{XY}\lesssim 10^{-20}$. This implies that 
\begin{equation}
    \mathcal{C}_2+\mathcal{C}_4/2 \lesssim 2\times 10^{-9}\text{ eV}^{-4}\,,
\end{equation}
which translates to a bound on the cutoff of $\Lambda \gtrsim 0.5$ keV.

Lorentz violation in the photon sector can also be probed with interferometers, because the speed of light is modified and depends on the direction of propagation. In the lab frame, we have the modification of the effective refractive index
$\bar{n}$~\cite{Kostelecky:2016kkn}: 
\begin{equation}
\label{eq:n}
    \delta \bar{n}=-\frac{1}{2}\hat{l}^j\hat{l}^k\Tilde{\kappa}^{jk}_{e-}+\frac{1}{2}\epsilon^{jkl}\hat{l}^j\Tilde{\kappa}^{kl}_{o+}+\Tilde{\kappa}_{\rm tr},
\end{equation}
where $\Tilde{\kappa}^{jk}_{e-}=-k^{jk}+\frac{1}{3}k_{ii}\delta^{jk}$,  $\Tilde{\kappa}^{jk}_{o+}=-\epsilon^{ijk}k_{0i}$, $\Tilde{\kappa}_{\rm tr}=\frac{2}{3}k_{00}$~\cite{Kostelecky:2002hh}, and $\hat{l}$ denotes the direction of the propagation. In the interferometer experiments, two beams of light are sent to two different directions $\hat{l}_1$ and $\hat{l}_2$ and compared when reflected back. The effect of the second term in Eq.~(\ref{eq:n}) cancels out in this two way measurement (since $\hat{l}$ changes sign on the way back). Then the difference between these two beams is only sensitive to $\Tilde{\kappa}^{jk}_{e-}$:
\begin{equation}
    \bar{n}_1-\bar{n}_2=\delta\bar{n}_1-\delta\bar{n}_2=-\frac{1}{2}(\hat{l}_1^j\hat{l}_1^k-\hat{l}_2^j\hat{l}_2^k)\Tilde{\kappa}^{jk}_{e-}.
\end{equation}
Again the Lorentz transformation from the lab frame to the SCCEF provides the (weaker) sensitivity of the experiment also to other LV parameters.

In~\cite{Kostelecky:2016kkn}, the data from Laser Interferometer Gravitational-Wave Observatory (LIGO) was analysed and the null LV signal~\footnote{In fact an anomaly with semi-annual frequency is found from the analysis. However this should not be interpreted as signal of LV since it is imcompatible with the constraints obtained from other frequencies.} gives strong bounds $|\Tilde{\kappa}^{XY}_{e-}|< 2.7 \times 10^{-22}$. Using Eq.~(\ref{eq:cmunu&kmunu}) and the relations under Eq.~(\ref{eq:n}), together with the positivity bound, this implies
\begin{equation}
     -2\times 10^{-10}\text{ eV}^{-4}\lesssim \mathcal{C}_4 \leq 0,
\end{equation}
or in terms of the cut-off, $\Lambda\gtrsim $ keV. 

\begin{figure}
     \centering
\includegraphics[width=.65\textwidth]{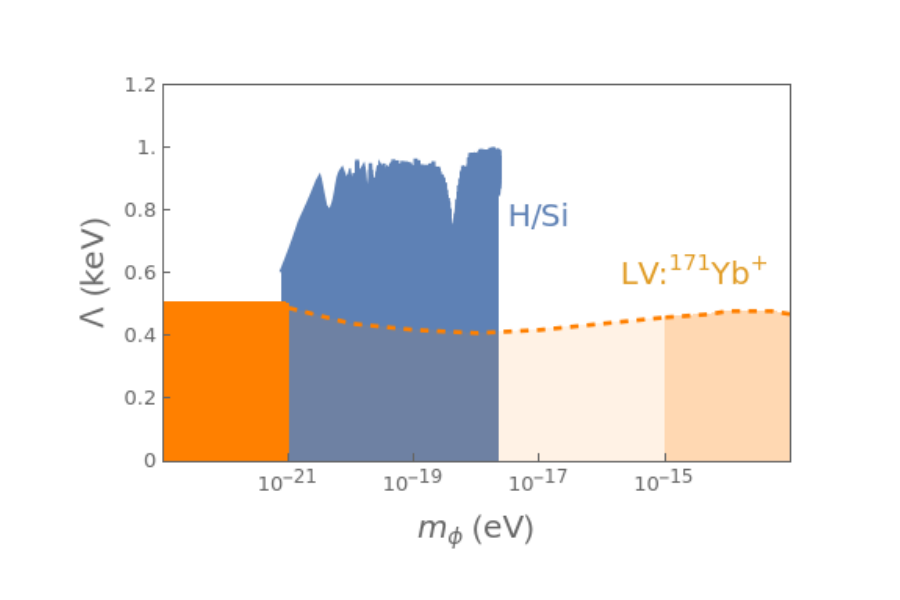}~     \caption{Bounds on the cut-off scale $\Lambda$ corresponding to $\mathcal{C}_2$. In the regime $m_\phi \lesssim 10^{-21} {\rm eV}$, the DM behaves as constant LV background and the bound can be obtained by direct interpretation of the bound from the experiment comparing two $^{171}$Yb$^+$ clocks~\cite{Sanner2019}. In the regime $10^{-21} {\rm eV} \lesssim m_\phi \lesssim 10^{-15} {\rm eV}$, the oscillating nature of $\phi$ field cannot be ignored. In the regime $ m_\phi \gtrsim 10^{-15} {\rm eV}$, the experiment is unable to resolve the fast oscillation of DM, but the signal will fluctuate at a time scale of the order of the DM coherence time. For the last two regimes, as indicated by the shaded region, a dedicated analysis of the experimental data needs to be done in the future to obtain the (precise) bound. With the blue region, we show that experiments comparing the frequencies of an H maser and silicon cavity~\cite{Kennedy2020} can be used to provide complementary bounds.}
     \label{fig:c2bounds}
 \end{figure}

Let us comment on the range of $m_\phi$ for which the bounds from the two experiments above apply. In the clock comparison experiment, the data is taken every roughly 2 seconds and covers a period of $T_e\approx  4 \times 10^6$ s. This means either $m_\phi \ll 10^{-21}$ eV so that the period of oscillation is much longer than $T_e$, or $m_\phi \gg 10^{-15}$ eV so that the oscillation is averaged out. For the analysis of data from LIGO, which covers a period of 16 months, the slow oscillation regime corresponds to $m_\phi \ll 10^{-22}$ eV. For the fast oscillation regime, we require $2\pi/m_\phi \ll 2L/c$, where $L\simeq 4$ km is the arm length of LIGO. This corresponds to $m_\phi \gg 10^{-11}$ eV. Away from these mass ranges, the oscillating nature of $\phi$ cannot be ignored when obtaining bounds from the experimental data. Besides these, for both experiments in the fast oscillation regime, the coherence time of ULDM $\tau_c \sim 1/(m_\phi v_{\rm DM}^2)$ is shorter than the total time period of the experiments. This means that the signal will fluctuate at the time scale of $\tau_c$ due to the stochastic nature of DM, and the bound could be affected by order one. We leave the thorough re-analysis of the experimental data, taking into account both the oscillating and stochastic nature of ULDM, to future work. Meanwhile, we notice that the experiments searching for oscillation of fundamental constants can provide complementary bounds. This is illustrated, as an example, in Fig.~\ref{fig:c2bounds}, where we show the bounds for $\mathcal{C}_2$ derived both from searching for LV and the H/Si comparison (by a direct translation of the bound on $\mathcal{C}_1 $, see section~\ref{sec:fund_oscillations} for discussion). Notice that the range of the bound from H/Si comparison is also limited by the total time period and the frequency of the data set.

\section{Astrophysical, cosmological and collider bounds} \label{sec:others}
In addition to terrestrial experiments, our theory can also leave signals in astrophysical and cosmological observations, warranting dedicated investigation that we leave for future work. However, we note the bounds derived from such observations often suffer from considerable uncertainties or rely on specific assumptions. For instance, the observation of high-energy cosmic rays could place strong constraints on Lorentz violation~\cite{Altschul:2021wrm,Klinkhamer:2007ak,HAWC:2019gui}. If the coefficient $\Tilde{\kappa}_{\rm tr}$ is positive, charged particles experience rapid energy loss through Cherenkov radiation when their energy exceeds a critical energy $E_c$ determined by $\Tilde{\kappa}_{\rm tr}$. Thus, observations of charged particles with energy $E_o$ that have traversed astronomical distances imply that $E_o < E_c$, leading to bounds on $\Tilde{\kappa}_{\rm tr}$. The specific bound for our case depends on the distribution as well as stochastic fluctuations of dark matter on the larger scales between the cosmic source and Earth.

There are additional bounds that do not assume $\phi$ as dark matter. The interaction between SM particles and light particles with masses $\lesssim 100$ MeV can be constrained by their effect on the rate of stellar cooling~\cite{Raffelt:1996wa}. As an example, consider the process of photon annihilation into a pair of $\phi$ induced by the operator $\mathcal{O}_3$. The cross section in the center of mass frame is given by $\sigma_{\gamma\gamma\rightarrow\phi\phi}=\mathcal{C}_3^2 \omega^6/32\pi$, where $\omega$ is the photon energy. In the thermal bath of photons, the energy loss rate per unit volume can be estimated as~\cite{Olive:2007aj} $\Gamma_{\gamma\gamma\rightarrow\phi\phi}=n_\gamma^2 \langle 2\omega \sigma_{\gamma\gamma\rightarrow\phi\phi}\rangle\approx 178 \mathcal{C}_3^2 T^{13}$. The constraint of the additional energy loss (per unit volume and time) in the supernovae burst SN1987A is about $\Gamma < 10^{-14} \text{ MeV}^5$, with the core temperature $T\approx$ 30 MeV~\cite{Raffelt:1996wa}. This gives the bound $\Lambda\sim \sqrt{4\pi}/\mathcal{C}_3^{1/4} \gtrsim 100$ GeV. The cooling of Horizontal Branch and Red Giant stars provide bounds in a similar way. With typical core temperature $T\approx 10$ keV, they give $\Lambda \gtrsim $ GeV, which are weaker then that from SN1987A, but still significantly stronger than the bounds derived from terrestrial experiments. However, it is noteworthy that all these cooling bounds can potentially be avoided through density (Chameleon) effects~\cite{        DeRocco:2020xdt,Bloch:2020uzh,Budnik:2020nwz}.

On the cosmological front, for $\phi$ being ultralight particles, they must decouple from the thermal bath prior to neutron decoupling to avoid changing the effective relativistic degrees of freedom and impacting the Big Bang Nucleosynthesis (BBN) process. This requirement leads to the condition $n \langle \sigma v \rangle < H$, where $n \sim T^3$ is the number density ($T$ being the temperature at decoupling), $\langle \sigma v \rangle \sim T^6/\Lambda^8$ is the averaged cross section, and $H\sim T^2/M_{\rm P}$ is the Hubble parameter ($M_{\rm P}$ being the Planck mass). Consequently, we obtain the constraint $\Lambda>(T^7M_P)^{1/8}$. Considering $T\sim$ MeV, we find $\Lambda \gtrsim$ GeV. Again, this stringent bound can potentially be circumvented through density effects.

Collider searches also offer valuable insights into our model. For instance, we can utilize data from the Large Electron-Positron Collider (LEP) on mono-photon events with significant missing energy to constrain the coupling of dark matter to electrons. The SM background in such cases is $e^+ e^- \rightarrow \gamma \nu \nu$, and it is found to be in agreement with measurements (see Table 6 in~\cite{L3:2003yon}). By imposing the requirement that the additional cross-section from $e^+ e^- \rightarrow \gamma \phi \phi$ is smaller than the uncertainty, we can constrain the couplings $\mathcal{C}_1$ and $\mathcal{C}_2$. We do not consider such bounds in this work, considering the possibility that $\phi$ could be a composite state, which should be replaced by more fundamental degrees of freedom at collider energies.

\section{Discussion and conclusions}\label{sec:dis&con}

In this paper we have studied an effective theory of scalar ultralight dark matter (ULDM) with quadratic couplings to the QED sector of the SM. Assuming a shift 
and $Z_2$ symmetry, the leading interactions between the ULDM and the SM 
arise from dimension-8 operators.  We calculated the one-loop RG mixing
of such operators and showed that to all orders they can be 
can be classified into two sectors, $j=0$ or $j=2$, 
without any RG mixing between them. Such operators can be associated with a variety of fundamental theories and in some cases may be the leading interactions between the ULDM and the SM -- as an explicit example we showed how they can be generated
by massive spin-2 graviton exchange. 
The ULDM background leads to Lorentz and Galilean violating effects, which we shall refer to collectively as Lorentz 
violating (LV).

\begin{table}
    \centering
    \small{\begin{tabular}{|c|c|c|c|c|}
        \hline
         & $\Lambda (\mathcal{C}_1)$ & $\Lambda (\mathcal{C}_3)$ &$\Lambda (\mathcal{C}_2)$ &$\Lambda (\mathcal{C}_4)$\\
        \hline
        OFC & keV  \small{(Yb/Cs~\cite{Kobayashi:2022vsf}, H/Si~\cite{Kennedy2020})}& 10 keV  \small{(Yb/Sr~\cite{Sherrill:2023zah})}& keV \small{(H/Si~\cite{Kennedy2020})}& \\
        \hline
        LV &  &  & 0.5 keV \small{(Yb~\cite{Sanner2019})}& keV \small{(LIGO~\cite{Kostelecky:2016kkn})}\\
        \hline
    \end{tabular}}
    \caption{The mass scales $\Lambda(\mathcal{C}_i)$, corresponding to bounds on the Wilson coefficients $\mathcal{C}_i$, that can be probed by the current experiments, either by  searching for oscillation of fundamental constants (OFC), or by looking for LV signals.}
    \label{tab:scales}
\end{table}

The sign of the Wilson coefficients ($\mathcal{C}$) in the $j=2$ sector is constrained by positivity bounds.  Regarding experimental tests of our theory, we considered two types of terrestrial experiments: those investigating the oscillation of fundamental constants and those searching for Lorentz-violating effects. We found that the first kind of experiment provides bounds on the Wilson coefficients of both sectors, while only the operators of $j=2$ sector are probed in the second kind of experiment. 
When the bounds are translated to a mass scale 
using $\mathcal{C}\sim 16\pi^2/\Lambda^4$, the current sensitivity of these two kinds of measurements only reaches the scale of $\Lambda \gtrsim \mathcal{O}(1-10)$ keV. See Table~\ref{tab:scales} for a summary of the scales probed by both types of experiments.
These bounds are found to be rather weak, because the effects of the dimension-8 operators in terrestrial experiments 
are suppressed by $\rho_{\rm DM}/\Lambda^4\approx (0.04 \rm{\, eV}/\Lambda)^4$. We note that this scaling is not unique to our model, but generally exists for other Lorentz-violating ULDM models. For instance, consider a dark photon $A'_\mu$. To obtain Galilean violation one must replace $\partial_\mu \phi \partial_\nu \phi \rightarrow A'_\mu A'_\nu$ in the operators 
discussed above. This would seem to generate dimension-6 operators, but there is an extra suppression of $m_A^2/\Lambda^2$ due to U(1) gauge symmetry. The dipole operator -- which is dimension 5 -- leads to magnetic-like DM background that interacts with electron spin, and generically cannot be probed in the clock comparison experiments.

The bounds obtained here could be improved by an order of magnitude with future clock experiments. Moreover, the potential existence of an ULDM solar halo~\cite{Banerjee:2019xuy,Banerjee:2019epw,Budker:2023sex}, leading to a bound DM density in the solar system approximately $10^4$ times larger than that of the standard halo model, would also elevate the bounds by a factor of 10.

Unlike the approach commonly discussed thus far in the literature (based on the SME ansatz), which is not based on a consistent effective theory, our EFT of ULDM gives a consistent description of Lorentz and Galilean symmetry violation. 
It leads to several other interesting implications, namely that we can consistently compare the bounds associated with Galilean symmetry violation to those associated with oscillation of fundamental constants and tests of equivalent-principle violation. Interestingly, enough, we find that these completely different three classes of experiments currently probe roughly the keV scale.  In addition, we have found that stronger but indirect bounds arise from astrophysical and cosmological observations, which probe the 0.1-100 GeV scale. However, these bounds are  model dependent, and can be avoided by 
density and temperature effects which may give rise to dramatic changes in the ULDM mass and coupling, thus inhibiting its production in stars or in the early universe.

\acknowledgments
The authors thank Marco Gorghetto, Wei Xue, Marianna Safronova and Hyungjin Kim for valuable discussions. The work of MJ is supported by a research grant from the Musk Foundation.
The work of GP is supported by grants from BSF-NSF, Friedrich Wilhelm Bessel research award of the Alexander von Humboldt Foundation, GIF, ISF, Minerva,
SABRA - Yeda-Sela - WRC Program, the Estate of Emile Mimran, and the Maurice and Vivienne Wohl Endowment. BP is grateful to the Weizmann Institute of Science for its kind hospitality and support through the SRITP and the Benoziyo Endowment Fund for the Advancement of Science.

\appendix
\section{Eliminating the $\overline{O}_1$ operator}
\label{appendix:operator elimination}

We can use the following field redefinition to absorb $\overline{O}_1$ into $O_1$
\begin{equation}
    \psi\longrightarrow\psi+\frac{\overline{c}_1}{m_\psi} \frac{1}{2}\partial_\mu\phi\partial^\mu\phi\psi\,,
\end{equation}
\begin{equation}
    \bar\psi\longrightarrow\bar\psi+\frac{\overline{c}_1}{m_\psi} \frac{1}{2}\partial_\mu\phi\partial^\mu\phi\bar\psi\,.
\end{equation}
Under the above redefinition,
\begin{equation}
 i\bar\psi\overset{\leftrightarrow}{\slashed{D}}\psi \longrightarrow i\bar\psi\overset{\leftrightarrow}{\slashed{D}}\psi +\frac{\overline{c}_1}{m_\psi}O_1\,,
 \end{equation}
 \begin{equation}
 -m_{\psi}\bar\psi\psi\longrightarrow -m_{\psi}\bar\psi\psi - \overline{c}_1\overline{O}_1\,,
 \end{equation}
where we ignored terms of higher dimension operators ($d>10$). Thus, the Lagrangian now becomes 
\begin{equation}
    \mathcal{L}\longrightarrow\mathcal{L}_{\text{QED}}+ \left(c_1+\frac{\overline{c}_1}{m_\psi}\right) O_1+\sum_{i=2}^4c_iO_i\,. 
\end{equation} 
We can now redefine the coefficient $c_1\longrightarrow c_1-\frac{\overline{c}_1}{m_\psi}$, which leads to the Lagrangian in Eq.~(\ref{eq:L_EFT}).

\section{RG flow calculation}
\label{appendix:RGflow}

\subsection{New vertices}

Apart from the QED vertex, our calculation involves six new vertices associated with the dimension-8 operators. Consider the expanded form of operators $O_1$ and $O_3$:

\begin{align*}
    O_1&=\partial_\mu\phi\partial^\mu\phi\, i \bar{\psi}\gamma^\nu \overset{\leftrightarrow}{D_\nu}\psi \nonumber\\ &=\underbrace{\frac{1}{2}\partial_\mu\phi\partial^\mu\phi \, i \bar{\psi}\gamma^\nu\partial_\nu\psi - \frac{1}{2} \partial_\mu\phi\partial^\mu\phi \, i (\partial_\nu\bar{\psi})\gamma^\nu\psi}_\text{V1} - \underbrace{ e \partial_\mu\phi\partial^\mu\phi \, A_\nu \bar{\psi}\gamma^\nu\psi \,, \nonumber}_\text{V3} \\ 
 O_3 &= \underbrace{\frac{1}{2}\partial_\mu\phi\partial^\mu\phi \,F_{\nu\rho} F^{\nu\rho} \,, \nonumber}_\text{V2}\\ 
\end{align*}

%\flushleft 
The terms V1, V2 and V3 correspond to the following vertices, 

\begin{figure}
    \centering
    \begin{equation*}
\begin{tikzpicture}
\begin{feynman}
\vertex (a);
\vertex [above left=of a] (i1);
\vertex [above right=of a] (f1);
\vertex [below left=of a] (i2);
\vertex [below right=of a] (f2);
\diagram* {
(i1) -- [scalar] (a) -- [scalar] (f1);
(i2) -- [fermion] (a) -- [fermion] (f2);
};
\end{feynman}
\end{tikzpicture}
\quad\quad
\begin{tikzpicture}
\begin{feynman}
\vertex (a);
\vertex [above left=of a] (i1);
\vertex [above right=of a] (f1);
\vertex [below left=of a] (i2);
\vertex [below right=of a] (f2);
\diagram* {
(i1) -- [scalar] (a) -- [scalar] (f1);
(i2) -- [photon] (a) -- [photon] (f2);
};
\end{feynman}
\end{tikzpicture}
\qquad
\begin{tikzpicture}
\begin{feynman}
\vertex (a);
\vertex [above left=of a] (i1);
\vertex [above right=of a] (f1);
\vertex [below left=of a] (i2);
\vertex [below right=of a] (f2);
\vertex [above=of a] (f3);
\diagram* {
(i1) -- [scalar] (a) -- [scalar] (f1);
(i2) -- [fermion] (a) -- [fermion] (f2);
(a) -- [photon] (f3);
};
\end{feynman}
\end{tikzpicture}
\end{equation*}
    \caption{New vertices}
    \label{New vertices}
\end{figure}
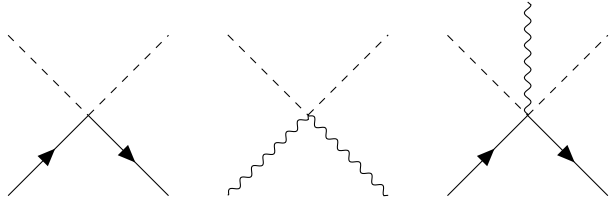

Operators $O_2$ and $O_4$ can be expanded similarly and they contain three more vertices that look identical to the above. 

\subsection{Diagrams contributing to operator mixing at one-loop order}

The RG equation for the Wilson coefficients follows from the structure of the operator renormalization matrix for $O_1$ -- $O_4$.  Operator renormalization in the $\overline{\hbox{MS}}$ scheme can be extracted from the UV divergent parts of photon-DM scattering and electron-DM scattering. There are several diagrams contributing to each of these processes, which are shown in Figures~\ref{fig:electron-DM} and~\ref{fig:photon-DM}.  
Evaluating these diagrams and requiring that the scattering amplitudes are UV finite after including wavefunction renormalization of the electron and photon fields leads to Eq.~(\ref{eq:RGE_CI}).

\subsubsection{First set of diagrams (electron-DM scattering)}

\begin{figure}
    \centering
    \begin{equation*}
\begin{tikzpicture}
\begin{feynman}
\vertex (a);
\vertex [above left=of a] (i1);
\vertex [above right=of a] (f1);
\vertex [below left=of a] (b);
\vertex [below right=of a] (c);
\vertex [below left=of b] (i2);
\vertex [below right=of c] (f2);
\diagram* {
(i1) -- [scalar] (a) -- [scalar] (f1);
(i2) -- [fermion] (b) -- [fermion] (a) -- [fermion] (c) -- [fermion] (f2);
(b) -- [photon] (c);
};
\end{feynman}
\end{tikzpicture}
\quad\quad
\begin{tikzpicture}
\begin{feynman}
\vertex (a);
\vertex [above left=of a] (i1);
\vertex [above right=of a] (f1);
\vertex [below left=of a] (b);
\vertex [below right=of a] (c);
\vertex [below left=of b] (i2);
\vertex [below right=of c] (f2);
\diagram* {
(i1) -- [scalar] (a) -- [scalar] (f1);
(i2) -- [fermion] (b) -- [fermion] (a) -- [fermion] (c) -- [fermion] (f2);
(a) -- [photon, half right, looseness = 0.8] (c);
};
\end{feynman}
\end{tikzpicture}
\quad\quad
\begin{tikzpicture}
\begin{feynman}
\vertex (a);
\vertex [above left=of a] (i1);
\vertex [above right=of a] (f1);
\vertex [below left=of a] (b);
\vertex [below right=of a] (c);
\vertex [below left=of b] (i2);
\vertex [below right=of c] (f2);
\diagram* {
(i1) -- [scalar] (a) -- [scalar] (f1);
(i2) -- [fermion] (b) -- [fermion] (a) -- [fermion] (c) -- [fermion] (f2);
(b) -- [photon, half right, looseness = 0.8] (a);
};
\end{feynman}
\end{tikzpicture}
\end{equation*}

\begin{equation*}
\begin{tikzpicture}
\begin{feynman}
\vertex (a);
\vertex [above left=of a] (i1);
\vertex [above right=of a] (f1);
\vertex [above=of a] (b);
\vertex [below=of a] (c);
\vertex [below left=of c] (i2);
\vertex [below right=of c] (f2);
\diagram* {
(i1) -- [scalar] (a) -- [scalar] (f1);
(i2) -- [fermion] (c) -- [fermion] (f2);
(a) -- [photon] (c);
(a) -- [fermion, half right, looseness = 0.75] (b) -- [fermion, half right, looseness=0.75] (a)
};
\end{feynman}
\end{tikzpicture}
% \quad\quad
% \begin{tikzpicture}
% \begin{feynman}
% \vertex (a);
% \vertex [above left=of a] (i1);
% \vertex [above right=of a] (f1);
% \vertex [below left=of a] (i2);
% \vertex [below right=of a] (f2);
% \vertex [above =of a] (b);
% \vertex [above =of b] (c);
% \diagram* {
% (i1) -- [scalar] (a) -- [scalar] (f1);
% (i2) -- [fermion] (a) -- [fermion] (f2);
% (a) -- [photon] (b);
% (b) -- [fermion, half right, looseness = 0.75] (c) -- [fermion, half right, looseness=0.75] (b)
% };
% \end{feynman}
% \end{tikzpicture}
\quad\quad
\begin{tikzpicture}
\begin{feynman}
\vertex (a);
\vertex [above left=of a] (i1);
\vertex [above right=of a] (f1);
\vertex [below left=of a] (b);
\vertex [below right=of a] (c);
\vertex [below left=of b] (i2);
\vertex [below right=of c] (f2);
\diagram* {
(i1) -- [scalar] (a) -- [scalar] (f1);
(i2) -- [fermion] (b) -- [fermion] (c) -- [fermion] (f2);
(a) -- [photon] (b);
(a) -- [photon] (c);
};
\end{feynman}
\end{tikzpicture}
% \quad\quad
% \begin{tikzpicture}
% \begin{feynman}
% \vertex (a);
% \vertex [below =of a] (d);
% \vertex [above left=of a] (i1);
% \vertex [above right=of a] (f1);
% \vertex [below left=of d] (i2);
% \vertex [below right=of d] (f2);
% \vertex [above =of a] (b);
% \vertex [above =of b] (c);
% \diagram* {
% (i1) -- [scalar] (a) -- [scalar] (f1);
% (i2) -- [fermion] (d) -- [fermion] (f2);
% (a) -- [photon] (b);
% (b) -- [fermion, half right, looseness = 0.75] (c) -- [fermion, half right, looseness=0.75] (b);
% (a) -- [photon] (d)
% };
% \end{feynman}
% \end{tikzpicture}
\quad\quad
\begin{tikzpicture}
\begin{feynman}
\vertex (a);
\vertex [above left=of a] (i1);
\vertex [above right=of a] (f1);
\vertex [below =of a] (b);
\vertex [below =of b] (c);
\vertex [below left=of c] (i2);
\vertex [below right=of c] (f2);
\diagram* {
(i1) -- [scalar] (a) -- [scalar] (f1);
(i2) -- [fermion] (c) -- [fermion] (f2);
(a) -- [fermion, half right, looseness = 0.75] (b) -- [fermion, half right, looseness = 0.75] (a);
(b) -- [photon] (c)
};
\end{feynman}
\end{tikzpicture}
\end{equation*}
    \caption{Electron-DM scattering diagrams}
    \label{fig:electron-DM}
\end{figure}
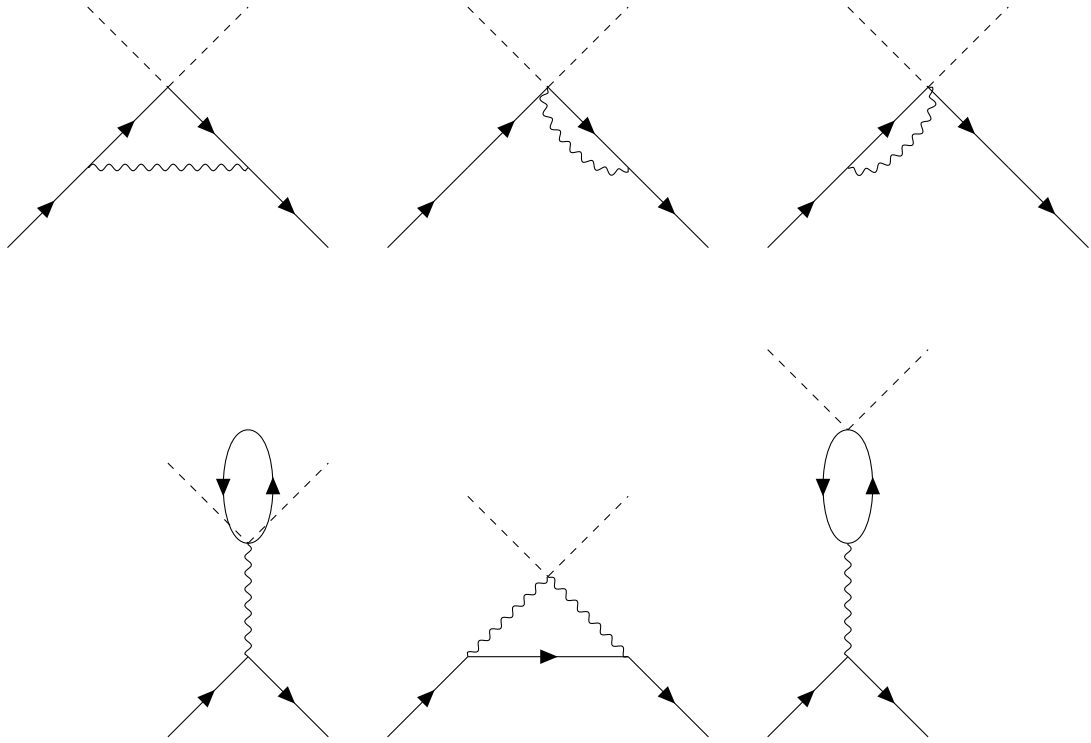

The relevant diagrams for electron-DM scattering are given in Figure \ref{fig:electron-DM}.

\subsubsection{Second set of diagrams (photon-DM scattering)}

\begin{figure}
    \centering
    \begin{equation*}
\begin{tikzpicture}
\begin{feynman}
\vertex (a);
\vertex [above left=of a] (i1);
\vertex [above right=of a] (f1);
\vertex [below left=of a] (b);
\vertex [below right=of a] (c);
\vertex [below left=of b] (i2);
\vertex [below right=of c] (f2);
\diagram* {
(i1) -- [scalar] (a) -- [scalar] (f1);
(i2) -- [photon] (b);
(c) -- [photon] (f2);
(a) -- [fermion] (b) -- [fermion] (c) -- [fermion] (a);
};
\end{feynman}
\end{tikzpicture}
\quad\quad
\begin{tikzpicture}
\begin{feynman}
\vertex (a);
\vertex [above left=of a] (i1);
\vertex [above right=of a] (f1);
\vertex [below left=of a] (b);
\vertex [below right=of a] (c);
\vertex [below left=of b] (i2);
\vertex [below right=of c] (f2);
\diagram* {
(i1) -- [scalar] (a) -- [scalar] (f1);
(i2) -- [photon] (b);
(c) -- [photon] (f2);
(a) -- [fermion] (c) -- [fermion] (b) -- [fermion] (a);
};
\end{feynman}
\end{tikzpicture}
\end{equation*}

\begin{equation*}
\begin{tikzpicture}
\begin{feynman}
\vertex (a);
\vertex [above left=of a] (i1);
\vertex [above right=of a] (f1);
\vertex [below left=of a] (b);
\vertex [below right=of a] (c);
\vertex [below left=of b] (i2);
\vertex [below right=of c] (f2);
\diagram* {
(i1) -- [scalar] (a) -- [scalar] (f1);
(i2) -- [photon] (b);
(a) -- [photon] (b);
(c) -- [photon] (f2);
(c) -- [fermion, half right, looseness = 0.8] (a) -- [fermion, half right, looseness = 0.8] (c);
};
\end{feynman}
\end{tikzpicture}
\quad\quad
\begin{tikzpicture}
\begin{feynman}
\vertex (a);
\vertex [above left=of a] (i1);
\vertex [above right=of a] (f1);
\vertex [below left=of a] (b);
\vertex [below right=of a] (c);
\vertex [below left=of b] (i2);
\vertex [below right=of c] (f2);
\diagram* {
(i1) -- [scalar] (a) -- [scalar] (f1);
(i2) -- [photon] (b);
(a) -- [photon] (c);
(c) -- [photon] (f2);
(b) -- [fermion, half right, looseness = 0.8] (a) -- [fermion, half right, looseness = 0.8] (b);
};
\end{feynman}
\end{tikzpicture}
\end{equation*}
    \caption{Photon-DM scattering diagrams}
    \label{fig:photon-DM}
\end{figure}
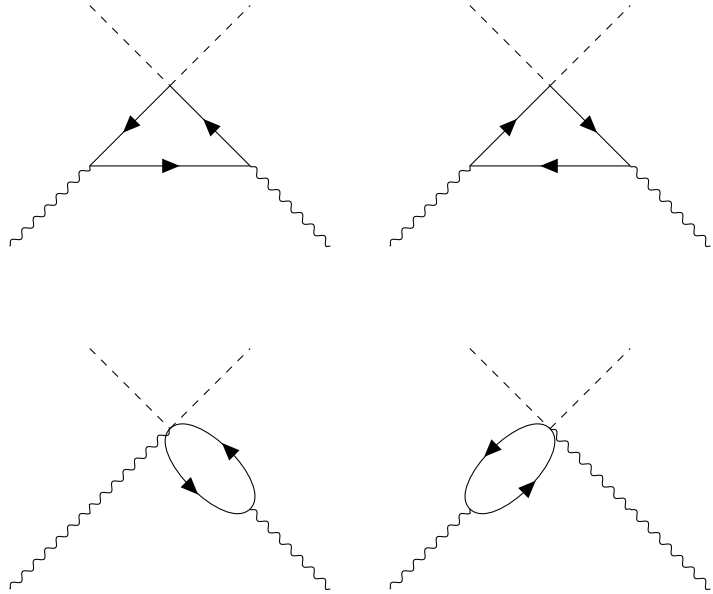

The relevant diagrams for photon-DM scattering are given in Figure \ref{fig:photon-DM}.

\section{Energy levels of the Hydrogen atom }\label{sec:H_levels}

In this section we give results for the dependence of the hydrogen atom energy levels on the Wilson coefficients $\mathcal{C}_{2}$ and $\mathcal{C}_{4}$, when the field $\phi$ takes on the background value in Eq.~(\ref{eq:wave}). To derive them,  we have obtained the total Hamiltonian for the electron-proton system interacting with the electromagnetic field in the non-relativistic limit using the standard techniques outlined in, for instance, \cite{Berestetskii:1982qgu}.
This involves taking the non-relativistic limit of the free Dirac
Lagrangian, and then calculating the interaction Hamiltonian from 
the Fourier transform of the Born-level electron-proton scattering amplitude in the same limit.   In presenting
the results, it is useful to introduce the quantities
\begin{align}
\tilde{c}\equiv \mathcal{C}_2 \rho_{\rm DM}\sin^2(m_\phi t) \, , \quad
\tilde{k}\equiv \mathcal{C}_4 \rho_{\rm DM} \sin^2(m_\phi t) \,.
\end{align}
Time derivatives acting on these objects are suppressed by $m_\phi$ and are thus neglected in what follows. 

Focusing on the terms relevant for the Rydberg energy, the spin-orbit interaction, and the spin-spin interaction, the resulting 
time-dependent Schrodinger equation takes the form
\begin{align}
%i\tilde{\partial}_t \psi  =
\left(1+\frac{3}{2}\tilde{c}\right) i\partial_t\psi =  \left(\tilde{H}_0 + \tilde{H}_{s-o}+\tilde{H}_{s-s} \right)\psi  \,. 
\end{align}
The Rydberg energy is obtained by setting the spin-orbit 
($\tilde{H}_{s-o}$)
and spin-spin ($\tilde{H}_{s-s}$) Hamiltonians to zero, and solving
the energy levels using the leading-order Hamiltonian
\begin{align}
\tilde{H}_0 = 
%- e \tilde{A}_0 - \left(1-\tilde{c}\right) \frac{\nabla^2}{2m_e} =
 -e\left(1+\frac{3}{2}\tilde{c}+\tilde{k} \right)A_0 - \left(1-\tilde{c}\right) \frac{\nabla^2}{2m_e}\, , 
\end{align}
with $A_0= e/(4\pi r)$ the Coulomb potential. Explicitly, one finds the eigenvalue equation
\begin{equation}
\left(-\frac{\tilde{\alpha}}{r}- \frac{\nabla^2}{2\tilde{m}_e}\right)\psi_{nlm}=\tilde{E}_n\psi_{nlm},
\end{equation}
where we have defined
\begin{align}
\tilde{\alpha} = \left(1+\tilde{k}\right)\alpha\, , \quad
\tilde{m}_e = \left(1+\frac{5}{2}\tilde{c}\right)m_e \,.
\end{align}
It follows that 
\begin{align}
\frac{\tilde{E}_n}{E_n} =  \frac{ \tilde{m}_e \tilde{\alpha}^2 }{m_e \alpha^2} = 1 + \frac{5}{2}\tilde{c}+2 \tilde{k}   
\end{align}
where $E_n$ are the standard energy levels of the hydrogen atom obtained by setting the Wilson coefficients set to zero.

We next turn to spin-orbit and spin-spin splittings, which are treated in perturbation theory. To this end, for a given value of $nlm$, we first write the energy level $E$ in presence of these perturbations as 
\begin{align}
\tilde{E} = \tilde{E}_n + \Delta\tilde{E}_{s-o} +   \Delta\tilde{E}_{s-s} \,.
\end{align}
For the spin-orbit Hamiltonian, we find 
\begin{align}
\tilde{H}_{s-o} = 
\left(1+\frac{3}{2}\tilde{c} \right) \frac{\tilde{\alpha}}{4\tilde{m}_e}\left(1+4\tilde{c} \right) \vec{\sigma_e} \cdot \left( \vec{\nabla} \frac{1}{r} \times i \vec{\nabla} \right)\,,  
\end{align}
where $\vec{\sigma}_e$ is the electron spin operator. From the fact that $\Delta E_{s-o}/E_n \propto \alpha^2$, it is easy to show that 
\begin{align}
\frac{\Delta \tilde{E}_{s-o}}{\tilde{E}_n} = \left(1+ 4 \tilde{c}+ 2 \tilde{k}\right)\frac{\Delta E_{s-o}}{E_n} \,.
\end{align}
Finally, for the spin-spin Hamiltonian, we find
\begin{align}
\label{eq:spin-spin}
\tilde{H}_{s-s} = 
%\frac{e}{2m_e}\vec{\sigma}_e \cdot \tilde{\vec{B}} = 
\frac{e}{2m_e}\left(1- \tilde{c}\right)\times \left(1-\tilde{k}\right)
\vec{\sigma}_e \cdot \vec{B} \, ,
\end{align}
where 
\begin{align}
\label{eq:Bfield}
\vec{B}=\frac{g_p e}{2m_p}\left[\frac{1}{4\pi r^3}[(3 \vec{\sigma_p}\cdot\hat{r})\hat{r}-\vec{\sigma}_p]+\frac{2}{3}\vec{\sigma}_p\delta^3(\vec{r})\right] \,.
\end{align}
Here $\vec{\sigma}_p$ and $m_p$ are the spin operator and mass of the proton, respectively, while $g_p$ is the $g$-factor parameterizing its magnetic moment. 
Proceeding in analogy with the 
spin-orbit term, we then find
\begin{align}
\frac{\Delta \tilde{E}_{\rm s-s}}{\tilde{E}_n}&=\left(1+\frac{5}{2}\tilde{c} \right)\frac{\Delta E_{\rm s-s}}{E_n} \, .
\end{align}

The dependence of the energy levels on the Wilson coefficients as given above  is a non-trivial result of the calculation. However, it  can be understood heuristically starting from the relevant terms in the standard NRQED Lagrangian (see~\cite{CASWELL1986437}):
\begin{align}
{\cal L}_{\rm NRQED} =
\psi_e^\dagger\left(iD^0 +\frac{\vec{D}^2}{2m_e} -\frac{e}{2m_e}
\vec{\sigma}\cdot \vec{B} - \frac{e}{8m_e^2} \vec{\sigma}\cdot 
\left[ i\vec{D}\times \vec{E} - \vec{E}\times i \vec{D}\right] \right)\psi_e \, , 
\end{align}
where $\psi_e$ is the non-relativistic electron field.
Note that, according to Eq.~(\ref{eq:O2&O4}), the effect of $\tilde{c}$ is to modify the time and spatial components of the covariant derivatives by
\begin{align}
\label{eq:c2rules}
D^0 \rightarrow \tilde{D}^0 =  (1+3\tilde{c}/2)D^0 \,, \quad 
D^i \rightarrow \tilde{D}^i = (1- \tilde{c}/2)D^i \, .
\end{align}
If we furthermore treat $A^\mu$ as an external field sourced
by the proton, with
\begin{align}
\label{eq:EandB}
\vec{E} = \vec{\nabla}A_0  \, , \quad \vec{B} = \vec{\nabla}\times \vec{A}
\end{align}
then we can obtain the remaining dependence on the Wilson coefficients through the modifications
\begin{align}
\vec{E} &  \to \tilde{\vec{E}} = \frac{1}{\epsilon}
\left(1 + \tilde{c} \right) \vec{E} \, ,   \, \nonumber \\
\vec{B} &  \to \tilde{\vec{B}} = \mu \left(1-\tilde{c} \right) \vec{B}  \, ,
\end{align}
where $\vec{E} = \vec{\nabla}e/(4\pi r)$ and $\vec{B}$  is given in 
Eq.~(\ref{eq:Bfield}). 
The dependence on $\tilde{c}$ in the above equation is obtained by applying replacements analogous to 
Eq.~(\ref{eq:c2rules}) on  the $A^0$, $\nabla^i$ and $A^i$ appearing  in Eq.~(\ref{eq:EandB}),  while the dependence on $\tilde{k}$ is reflected in the change in the permittivity and permeability in the Maxwell Lagrangian as given in 
Eq.~(\ref{eq:eps_and_mu}).

\bibliographystyle{JHEP}
\bibliography{phi}

\end{document}